# Local asymmetry in spatial interactions: A generalized slide-vector approach

Bin Liu[ab1], Zhaoya Gong[ab1*], Jean-Claude Thill[c]

[a]*School of Urban Planning & Design, Peking University Shenzhen Graduate School, Shenzhen, Guangdong, China;*
[b]*Key Laboratory of Earth Surface System and Human-Earth Relations of Ministry of Natural Resources of China, Peking University Shenzhen Graduate School, Shenzhen, Guangdong, China;*
[c]*Department of Earth, Environmental and Geographical Sciences and School of Data Science, University of North Carolina at Charlotte, NC, USA;*

[1]*These authors contribute equally to this paper*
*Corresponding author (Email: z.gong@pku.edu.cn)*

# Local asymmetry in spatial interactions: A generalized slide-vector approach

The conceptualization of space is crucial for comprehending the processes that shape geographic phenomena. Functional space exhibits asymmetric spatial separations, which deviate from the symmetry axiom of metric space commonly adopted as a representation of the geographical space. However, existing literature has paid scant attention to the issue of asymmetry of spatial separation. Technically, spatial models and analysis methods grounded in a Euclidean representation of the geographical space have their capability to handle the functional space of geographical phenomena restricted by the inherency of the symmetry axiom. In this study, we aim at differentiating and characterizing the spatial dependency and heterogeneity of asymmetric spatial separations. Specifically, we propose a local slide-vector model based on spatially constrained multi-dimensional unfolding. The model takes account of spatial dependency and heterogeneity of asymmetry and can capture local asymmetric structures of spatial separations. Furthermore, we examine the dynamics of local asymmetric structures and introduce a potential field method to infer inter-regional asymmetries. To demonstrate the validity of our approach, we apply it to study the spatial separations derived from U.S. interstate migration data. Our approach sheds light on the distortion of geographic space from the perspective of migrants' relocation preferences and improves the understanding of domestic human migration patterns.

# 1 Introduction

Conceptualization of space drives the understanding of the spatially interactive processes that shape geographic phenomena (Mazúr and Urbánek 1983; Thill 2011). Functional spaces are closely coupled with human cognition of physical space and with human experience of places. They encapsulate the various functional relationships between places embedded in the physical space (Papadakis, Resch, and Blaschke 2020; Li, Ma, and Wilson 2021). In essence, they can be understood as relational spaces constructed by the interactions among individuals or organizational entities (Doel 2007; Jones 2009; Malpas 2012; Bergmann and O'Sullivan 2018).

We contend that functional spaces exhibit asymmetric spatial separations. This marks a significant departure from the symmetry axiom of metric space as a customary representation of the geographical space (Couclelis 1999, Christakos 2000). It is common to observe imbalances in bilateral international trade and migration flows; these may stem from the asymmetric geographical costs or barriers between paired origins and destinations (Codol et al. 1989; Selmer, Chiu, and Shenkar 2007). However, given that most existing spatial models and analysis methods rely on a Euclidean representation of the geographical space, they have limited capability to represent and analyze the functional spaces of geographic phenomena exhibiting asymmetric relations (Midler 1982; Couclelis 1999; Goodchild et al. 2007; Paasi 2011; Allard et al. 2024).

Existing literature has paid insufficient attention to the asymmetry of spatial separation. This situation can be attributed to at least two main points. First, most studies hold that asymmetry has a negligible effect when studying spatial separations, as it is common that asymmetric proportions are dramatically smaller than symmetric proportions (Liu et al. 2014; Wesolowski et al. 2015). Consequently, asymmetries in the data are usually treated as noise or errors, and hence removed by simply averaging the asymmetric distances in the proximity matrix. However, relative insignificance does not mean that asymmetry should be discarded as a nuisance. Second, representing asymmetry in high-dimensional spatial interaction data is simply nontrivial. Tobler (1975) pioneered modeling asymmetry in human migration through a vector-field-based representation analogous to the 'wind field' that models the asymmetric migration flows between locations. A few recent studies have adopted similar vector-field approaches to modeling asymmetry in spatial interactions, including regional trade flow (Broitman et al. 2021), human mobility (Yang et al. 2023), and traffic flow (Ji et al. 2020; Ji et al. 2022). On the other hand, alternative approaches have recognized asymmetry as an important property of cognized spatial separation in spatial interactions and have tried to derive asymmetric functional distances from inverted spatial interaction models (Plane 1984).

Notwithstanding the few studies above, the majority of research has treated the asymmetry in spatial interaction data as a deviation from regular symmetric patterns and has seldom modeled asymmetry as an inherent property of functional spaces, which are themselves reconstructed from spatial interactions (Ahmed and Miller 2007). Functional space can represent spatial dependence and heterogeneity more faithfully via functional distance than absolute space with Euclidean distance when modeling spatial interaction phenomena (Miller 2004). This is because functional space relaxes the restriction of metric space's geometric axioms, among which asymmetry is a critical one. Asymmetry is one of intrinsic properties of spatial interaction phenomena (Tobler 1979; Liu et al. 2024). Meanwhile, asymmetry of spatial interaction naturally intertwines with its two other important characteristics, spatial dependence and heterogeneity, and may exhibit global and local structures. However, almost no study has focused on the spatial dependence and spatial heterogeneity of asymmetry in spatial interactions and it remains to be determined whether asymmetric spatial interaction exhibits global and/or local structures.

To close this research gap, this paper aims at characterizing and differentiating the global versus local structures of asymmetric spatial separations. Specifically, we seek to answer the following three questions:

Q1: What are local versus global asymmetries in functional spaces constructed from spatial separations? How can local asymmetry in functional spaces be modeled by accounting for the spatial dependency and heterogeneity of asymmetries?

Q2: Do local asymmetric structures in migration spaces exhibit dynamics over time? How can the dynamic structures of local asymmetries be modeled to reveal migration trends within local regions?

Q3: What are the asymmetric migration patterns between local regions? How to infer inter-regional asymmetries based on asymmetric structures of local regions?

To answer these questions, we propose a local slide-vector model that is capable of considering the spatial dependency and spatial heterogeneity of asymmetries, based on spatially constrained multi-dimensional unfolding. Furthermore, we introduce a potential field approach to representing the uncovered asymmetric structures of functional spaces. To demonstrate the validity of our approach, we apply it to the spatial separations derived from U.S. bilateral interstate migration data. This allows us to reconstruct the migration space and visualize the migration-scape, revealing hitherto undisclosed asymmetric structures. Our approach sheds light on the distortion of the geographic space from the perspective of migration flows and facilitates an improved understanding of domestic migration patterns.

The rest of the paper is arranged as follows: In Section 2, we systematically review relevant literature and point out the problems of existing methods. In Section 3, we will provide a detailed introduction to the slide-vector model proposed in this paper. In Section 4, we introduced the data, study area, and experimental setup. In Section 5, we will demonstrate the superiority of our method and showcase the asymmetric spatial distortions that differ from static geographic space. Finally, in Section 6, we will draw conclusions, highlight the contribution, and point to future research priorities.

## 2  Literature review

### *2.1 The asymmetric nature of functional spaces*

The conceptualization of space is of paramount importance to comprehend the processes that shape spatial interaction phenomena. With the waning of the friction of distance brought by advances of information and communication technologies and transportation means, it has garnered renewed attention centered on the transformation of space concept from a static container (absolute space) to a relational or functional one constructed through relations or relatedness between entities that collectively constitute a space (Goodchild et al. 2007; Thill 2011; Shaw and Sui 2020). As spatial interaction phenomena commonly violate metric space axioms, such as symmetry and triangular inequality, the absolute space approach rooted in the Euclidean metric is not an appropriate representation for them (Sheppard 2006; Ahmed and Miller 2007; Miller and Bridwell 2009). Conversely, the functional space approach may employ a non-Euclidean semi-metric or quasi-metric to encapsulate relational or functional proximity, demonstrating greater analytical and visual efficacy than Euclidean metric (Bunge 1966; Midler 1982; Miller 2000; Tobler 1993). Fittingly, new modeling tools have been developed to estimate these non-Euclidean space from observed spatial interactions (Ahmed and Miller 2007; Bunge 1966; Cliff and Haggett 1998).

Notably, when modeling spatial interaction phenomena, functional space represents spatial dependence and spatial heterogeneity more faithfully via functional distance than absolute space fitted with Euclidean distance possibly can. For example, long-distance interactions and spacetime compression seem to violate spatial dependence from the standpoint of Euclidean distance, but they are actually conforming with it from a perspective of functional distance (e.g., travel time or interaction frequency), as functional space operationalizes "proximity" through functional distance instead of Euclidean coordinates (Miller 2004). Additionally, heterogeneity of functional spaces has been well recognized. Recent advances in spatial interaction modeling demonstrate that embedding mobility networks through deep learning and dimensionality reduction reveal heterogeneous clustering patterns in latent spaces (Crivellari and Beinat 2019; Damiani et al. 2020; Fan, Yang, and Mostafavi 2024). These clusters, corresponding to distinct community structures in mobility, provide empirical evidence for the heterogeneity of functional spaces.

More importantly, we argue that functional distance is better suited to capture the dependence and heterogeneity of spatial interactions as it is not impeded by the metric space's geometric axioms, among which asymmetry is a crucial one. Asymmetry is an inherent property of spatial interaction phenomena, after all purely symmetric spatial interaction processes have been controlled for. As discussed by Tobler (1979), spatial interaction tables can be decomposed into symmetric and asymmetric components, with the symmetric part representing the essential latency of exchanges and the asymmetric part revealing the predominant flow directionality stemming from disparities across the geographic space (Constantine and Gower 1982). Moreover, empirical studies have demonstrated that incorporating asymmetry accounts for more nuanced spatial proximity and enhances the understanding of spatially dependent and heterogeneous interaction patterns. For instance, it has been found that people choose the 'least effort' route based on convenience rather than physical length, where the former is commonly asymmetric while the latter is always symmetric (Bongiorno et al. 2021; Ben-Elia et al. 2013; MacEachren 1980). Plane (1984) demonstrated that functional spaces constructed from interstate migration flows exhibit greater explanatory power than geographic spaces in understanding the heterogeneity of destination choices made by migrants.

As an inherent property of spatial interaction, asymmetry inevitably intertwines with its two other characteristics, namely spatial dependence and heterogeneity. On one hand, asymmetry can more effectively reflect the functional proximity between locations in spatial interaction phenomena, thereby representing spatial dependence and heterogeneity more faithfully. On the other hand, asymmetry of functional proximity itself naturally becomes spatially dependent and heterogeneous; e.g., traffic congestion propagates to the connected roads due to the outbound commuting towards the city center while, in the opposite direction, those roads are clear of traffic. However, almost no study has focused on the spatial dependence and spatial heterogeneity of asymmetry in spatial interactions and none has examined whether it exhibits global and/or local structures.

## *2.2 Modeling asymmetry in functional spaces*

A typical approach to modeling functional spaces is through transformations of functional distances (Tobler 1961). Methods such as multidimensional scaling (MDS) were employed to reflect the distorted patterns of space (Marchand 1973; Ewing 1974). However, ordinary MDS techniques assume that distances between objects are symmetric. These studies typically consider asymmetry as noise or attributed to errors, and attempt to eliminate the asymmetry by averaging the corresponding elements in the bidirectional matrix (Rothkopf 1957; Shepard 1963). For example, Ahmed and Miller (2007) compared the symmetric and asymmetric solutions of MDS for travel time data and concluded that the symmetric solution was sufficiently good and the consideration of asymmetry was not needed. Some other studies ignored the asymmetric part outright for the sake of data simplification, on the premise that the undirected representation encoded most of the interactions (Johnson 2020). This assumption is not always neutral, particularly when the asymmetry of spatial interactions is structural, such as in exchange flows, migration, or social network interactions, where asymmetry may be intrinsic to the spatial interaction phenomena under study.

Both theoretical and empirical investigations into spatial interactions pay surprisingly little attention to structural asymmetry in modeling functional spaces, not even mentioning its global and local patterns. A few studies in the field of human geography recognized these asymmetric structures in spatial interaction phenomena. Plane (1984) derived the asymmetric cognitive distances by inverted doubly-constrained spatial interaction models, allowing for the construction of "migration spaces" that reflect origin-specific and destination-specific perspectives. These asymmetric structures were visualized individually using linear cartograms centered on each location, comparing how migrants from or to a specific location perceive functional separation. However, the analysis was confined to individual origin–destination perspectives and visual in nature, therefore do not support analysis of asymmetries between locations. In other words, the potential spatial dependency and heterogeneity among local asymmetries across locations remained unexplored, limiting our understanding of the broader regional and global organization of asymmetry.

An alternative approach aimed to model asymmetry directly by fitting a vector field to spatial interaction data such as migration flows, analogous to a wind field or current field (Tobler 1975; Mazzoli et al. 2019). In this approach grounded in the field theory in physics, asymmetries are represented such that interactions in the direction of vectors in the field are more likely to occur, while those against those vectors face greater spatial resistance. Structural asymmetries are further transformed into a potential field from the vector field, indicating locations with high or low migration potentials as hubs of inward or outward migration. This field approach signifies an attempt to represent and visualize the asymmetric structures of spatial interactions with a reference to the geographic space. However, it lacks the reconstruction of relevant functional spaces pertaining to the spatial interaction phenomena, which must be built on derived functional metrics or distances.

Outside the field of geography, asymmetric multidimensional scaling (AMDS) was proposed to model asymmetric dissimilarity data. For example, Young's ALSCAL (Young, Takane, and Lewyckyj 1978) and Chino's ASYMSCAL (Chino 1978) aim to produce asymmetric multidimensional configurations that can be interpreted visually. The radius-distance model (Okada and Imaizumi 1987) and the slide-vector model (Zielman and Heiser 1993) superimpose asymmetric information onto symmetric MDS plots. Kruskal and Wish (1978) suggested splitting the matrix into n rows, performing a separate monotonic regression for each row, and then combining the stress values into a single statistic; this procedure is known as multi-dimensional unfolding (MDU). However, these methods have primarily been developed for psychometrics and do not feature a rich spatial perspective, especially with regards to the local elements of asymmetries, namely their spatial dependency and heterogeneity, as their primary function is to generalize and reduce dimensionality.

In sum, existing approaches are inadequate to effectively represent and quantify the asymmetric structures in modeling functional spaces of spatial interactions. Moreover, these approaches fall short in balancing the consideration of global and local structures of asymmetries in their models, which hinders a systematic understanding of asymmetry of functional spaces.

## 3 Methodology

To bridge the gaps in theory and methodology identified above, we propose a novel approach grounded in a non-Euclidean spatial separation metric derived from spatial interaction to represent the asymmetry of functional space. This approach addresses the challenge of balancing global and local features by enabling a systematic understanding of structural asymmetry patterns. Furthermore, we introduce a potential field model that encapsulates spatial resistance, allowing for the representation of the asymmetrical terrain in a three-dimensional space.

### *3.1 Slide-vector model*

The slide-vector model (SVM) is based on multidimensional unfolding (MDU) that inputs an asymmetric origin-destination spatial separation matrix of $n$ locations among which each location's origin and destination roles are treated as different entities. It outputs a configuration of the set of entities, denoted as $\{X, Y\}$ consisting of $n$ locations' origin and destination entities X and Y with their coordinates $\{(x_i, y_j) | i, j = 1 \cdots, n\}$. The objective is to approximate the pairwise Euclidean distances $\{d_{ij}(X, Y)\}$ as closely as possible to their given spatial separations $\{\delta_{ij} | \delta_{ij} \neq \delta_{ji}\}$. A standard MDU is formulated as a stress minimization problem (Kruskal 1964):

$$stress = min\left\{\sum_{i=1}^{n}\sum_{j=1}^{n}\left(\delta_{ij} - d_{ij}(X,Y)\right)^2\right\} \quad (1)$$

With the fitted Euclidean distance expressed as:

$$d_{ij}(X,Y) = \sqrt{\sum_{k=1}^{K}(x_{ik} - y_{jk})^2} \quad (2)$$

where K represents a K-dimensional Euclidean space, both X and Y are $n \times k$ matrices, representing the origin and destination configurations, respectively.

By enforcing a global constraint vector Z, the original SVM model (Zielman and Heiser 1993) results in a solution configuration where there is a uniform shift between the corresponding positions of origin and destination entities. This global shift, represented by Z with a fixed magnitude and direction, captures the global asymmetry.

Formally, inserting the SVM constraint $y_{jk} = x_{jk} - z_k$ into Equation (2) yields:

$$d_{ij}(X,Z) = \sqrt{\sum_{k=1}^{K}(x_{ik} - x_{jk} + z_k)^2} \quad (3)$$

Figure 1(a) illustrates geometrically how the slide-vector Z affects the directional distances between locations P and Q in a two-dimensional space. A location's origin (O) and destination (D) positions are marked by circle and triangle, respectively. The O positions of P and Q are designated as vectors *p* and *q* with reference to the origin of the coordinate system. Their directional distances are symmetric, as (*p* - *q*) and (*q* - *p*) have equal length but opposite directions. With a slide-vector Z (marked by red arrow) constraint, a location's D position becomes a result of translating its O position along the direction and with the length of $Z$. By doing this, the distance from P's O position to Q's D position and the one from Q's O position to P's D position, $q - p + z$ and $p - q + z$ respectively, are no longer the same and hence become asymmetric. This translational operation is geometrically significant as it quantifies the directional difference between the entities beyond mere symmetrical proximity.

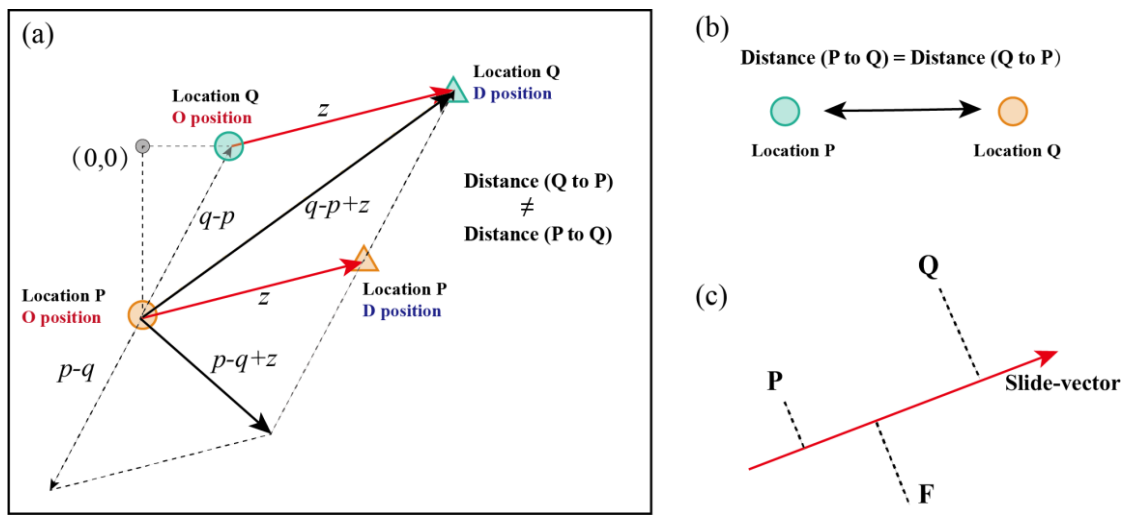

Figure 1. Definition of spatial separation and functional asymmetry distance: (a) geometry of SVM and functional asymmetric spatial separation (modified from Zielman and Heiser 1993); (b) traditional geography of symmetric distance between two locations; (c) joint representation of SVM.

Figure 1(c) visually explains how the slide-vector Z and embedding points jointly represent a global asymmetry. The arrow represents the direction of the slide vector, which indicates the asymmetry that it is more costly traveling along this direction than in the opposite direction. Assume that there is a real axis along the direction of the slide-vector with larger values towards the arrow side. The dashed lines indicate projections of locations onto the real axis. Given the above asymmetry, location Q with the largest projected value on the real axis is more costly to be reached from other locations, compared to traveling from location Q to other locations like P or F. In other words, locations with higher projected values have more importance as origins than destinations; otherwise, they have more importance as destinations than origins. On the other hand, distances between projected points can be viewed as measures of similarity, with location P more similar to F than to Q. The rationale behind the model is that symmetric distances can be reviewed as climbing a mountain, where the effort varies between uphill and downhill movements, yet the distance stays constant (also depicted in Figure 3).

### *3.2 Generalizing slide-vector model*

*Local versus global slide-vector model*

The original SVM is limited in that it only preserves the global asymmetry characteristics while obscuring the variation of local asymmetries due to the uniform constraint on all vectors. Here, we generalize the SVM by introducing region-specific constraints, enabling a local version of the SVM to capture local asymmetric structures by accounting for the spatial dependency and heterogeneity of asymmetries.

We assume that $n$ locations can be grouped into $m$ local regions, given the existence of certain spatial dependency between their asymmetries. The total number of regions satisfies $m \leq n$; the upper bound corresponds to the extreme case where each location forms its own region, indicating a completely spatial independency of asymmetries. Within each region $R_l$ ($l = 1,2,\cdots,m$), locations share a locally defined slide-vector $Z_l$. Inserting the constraint $y_{jk} = x_{ik} - z_{lk}$ in Equation (2) yields the definition of the distance between locations $i, j \in R_l$:

$$d_{ij}(X, Z_l) = \sqrt{\sum_{k=1}^{K}(x_{ik} - y_{jk})^2} = \sqrt{\sum_{k=1}^{K}(x_{ik} - x_{jk} + z_{lk})^2} \tag{4}$$

These local slide-vectors can be similarly solved by minimizing the stress function:

$$stress = \min\left\{\sum_{l=1}^{M}\sum_{i,j \in R_l}\left(\delta_{ij} - d_{ij}(X, Z_l)\right)^2 + \sum_{\substack{i \in R_p, j \in R_q \\ p \neq q}}\left(\delta_{ij} - d_{ij}(X, Y)\right)^2\right\} \tag{5}$$

where the second item is for distances between locations at different local regions.

To formalize Equation (4) in matrix form, let $Z^* \in \mathbb{R}^{m \times K}$ stack the local slide-vectors $Z^* = [Z_1, Z_2, Z_3, \ldots, Z_m]^T$. Our local slide-vector based constraints are then given by $Y = X - AZ^*$. Stacking the row (origin) and column (destination) coordinates underneath each other gives:

$$\begin{bmatrix} X \\ Y \end{bmatrix} = \begin{bmatrix} X \\ X - AZ^* \end{bmatrix} = \begin{bmatrix} I & 0 \\ I & -A \end{bmatrix}\begin{bmatrix} X \\ Z^* \end{bmatrix} = E\begin{bmatrix} X \\ Z^* \end{bmatrix} \tag{6}$$

where $A \in \mathbb{R}^{n \times m}$ is a location assignment matrix with $a_{il} = 1$ if location $i$ belongs to region $R_l$ and 0 otherwise, which enforces a one-hot structure to ensure locations in the same region share the same local slide-vector. The method for obtaining $A$ is detailed in Section 3.2.2. The composite matrix $E$ is used as a configuration matrix for linear constraints.

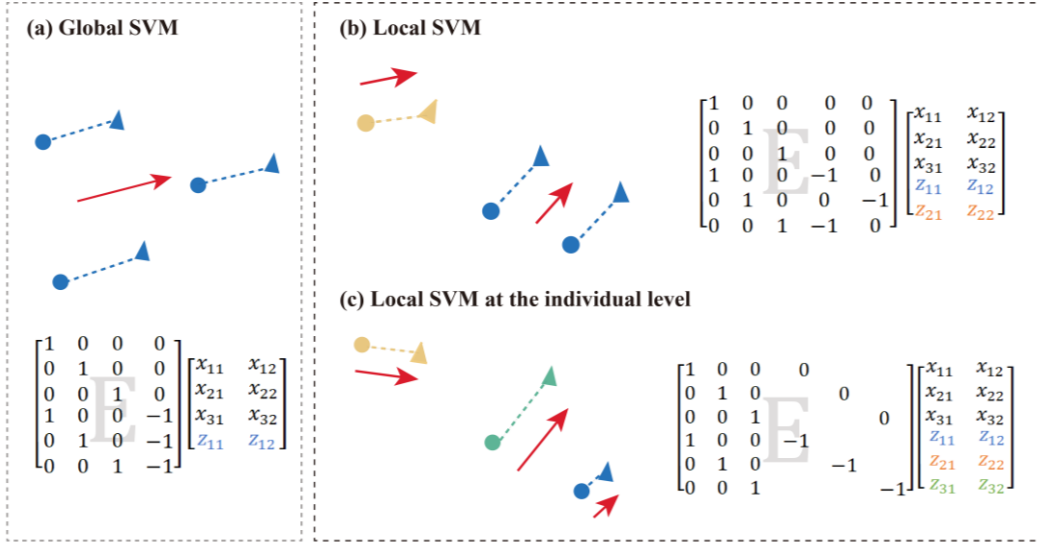

Figure 2. Illustrative examples of global versus local slide-vectors. Circular dots represent origin positions, triangular dots represent destination positions, and red arrows indicate slide-vectors.

Figure 2 illustrates different SVMs involving three locations. The constraints for the local SVM are defined in Equation (7),

$$E\begin{bmatrix}X\\Z^*\end{bmatrix}=\begin{bmatrix}1&0&0&0&0\\0&1&0&0&0\\0&0&1&0&0\\1&0&0&-1&0\\0&1&0&0&-1\\0&0&1&-1&0\end{bmatrix}\begin{bmatrix}x_{11}&x_{12}\\x_{21}&x_{22}\\x_{31}&x_{32}\\z_{11}&z_{12}\\z_{21}&z_{22}\end{bmatrix}=\begin{bmatrix}x_{11}&x_{12}\\x_{21}&x_{22}\\x_{31}&x_{32}\\x_{11}-z_{11}&x_{12}-z_{12}\\x_{21}-z_{21}&x_{22}-z_{22}\\x_{31}-z_{11}&x_{32}-z_{12}\end{bmatrix} \quad (7)$$

where $(x_{11}, x_{12})$ represents the configuration of the first location as the origin position, $(x_{11} - z_{11}, x_{12} - z_{12}) = (y_{11}, y_{12})$ represents the configuration of the first location's destination position belongs to region $R_1$ under the local SVM. Configurations belonging to different local clusters are subjected to distinct local slide-vectors under the linear constraint of $E$.

As shown in Figure 2(a) and (b), unlike the global SVM which captures the globally uniform asymmetry, the local SVM reveals region-specific asymmetric structures based on the spatial dependencies and heterogeneity of asymmetries. In the example, two locations share the same local asymmetry, while the third exhibits a distinct pattern. Notably, local SVM at the individual level is an extreme case of local SVM, as illustrated in Figure 2(c), where each location has a unique asymmetry pattern subject to no spatial dependency.

*Identifying dependency and heterogeneity of local asymmetries*

To operationalize local SVM, the first step is to define the location assignment matrix $A$, which is formally defined in Equation (6). Constructing $A$ requires identifying a set of regions $\{R_l\}_{l=1}^{m \leq n}$, where each region contains locations exhibiting similar local asymmetry pattens. These regions are constructed based on the finest-level asymmetry patterns obtained from the local SVM at the individual level.

Specifically, we first define the measure for evaluating similarity of asymmetries. Using the individual spatially separated O-D vectors from the individual-level SVM, which capture the asymmetry of single locations, the identification of similarity between individual slide-vectors relies on two primary criteria: 1) the similarity in length and direction; 2) the spatial proximity of the vectors. We adopt the modified distance measure developed by Tao and Thill (2016) for spatial flow, as shown in Equation (8):

$$FVD(u,v) = \frac{\sqrt{d_O(u,v)^2 + d_D(u,v)^2}}{cosin(u,v)} \quad (8)$$

where $FVD(u,v)$ represents the vector distance between vectors $u$ and $v$; $d_O(u,v)^2, d_D(u,v)^2$ measures the similarity of distances between their starting and ending points, and $cosin(u,v)$ measures the directional similarity between them.

Next, we employ cluster detection to examine the spatial dependency and heterogeneity among individual slide-vectors (Getis and Franklin 2010; Tao and Thill 2016). The cluster detection captures spatial dependence through the proximity and similarity of events within clusters, while spatial heterogeneity is revealed by the dissimilarity between clusters.

We adopt the modified distance measure in Equation (8) as a dissimilarity metric between clustering elements. HDBSCAN, a hierarchical density-based clustering algorithm (McInnes, Healy, and Astels 2017), is chosen here for two primary reasons: 1) it can adaptively determine the number of clusters based on the data without requiring complex parameter tuning; 2) it is capable of identifying and filtering out noise vectors rather than forcefully assigning them to clusters.

In the context of the local SVM, each cluster detected by HDBSCAN corresponds to a local region $R_l$. Locations identified as noise vectors are treated as singleton regions—containing only one location—and thus retain their individual constraints. Together, these cluster and singleton regions define the full set $\{R_l\}_{l=1}^{m \leq n}$ used to construct the location assignment matrix $A$, enabling the model to accommodate both spatial dependency and heterogeneity of local asymmetries.

### *3.3 Representation of asymmetric spaces as potential fields*

We consider locations in a given vector cluster to have the same local asymmetric pattern, where the local vectors of the unified structure are isotropic, equidistant and parallel after fitting through the local slide-vector model.

In Figure 3, the asymmetry of the O-D and D-O distances between A and B can be intuitively understood. The net flow is from A to B, as the spatial flow from A to B exceeds that from B to A. In our slide-vector, the spatial separation from B to A is greater than that from A to B. This asymmetry can metaphorically be viewed as climbing a mountain, where the spatial resistance from B to A has a net balance over the reverse direction, and the direction of the slide-vector represents the direction of larger spatial costs.

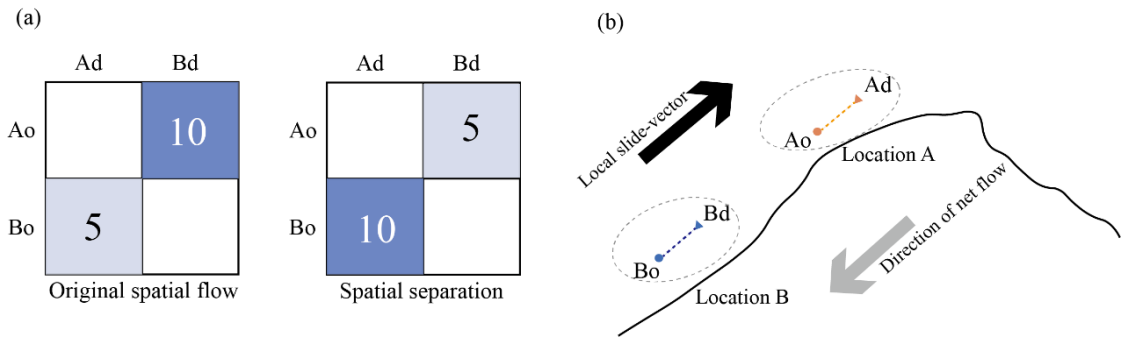

Figure 3. Metaphor of SVM

Under this metaphor, one can consider a representation of asymmetric distances between sites on a hill slope. We first assume the existence of a continuous latent vector field $\vec{c}(x, y)$ from which the fitted local slide-vectors are sampled; then, the latent vector field of local asymmetry can be interpolated.

Two additional assumptions are required for vector field interpolation. First, the locality of asymmetry serves as the basis for interpolation. The local slide-vector represents the asymmetry of a specific region, implying shared asymmetries within local neighborhoods. Individual-level slide-vectors that do not fall into any cluster are excluded from the interpolation process, as they solely represent the asymmetry of a single location and do not necessarily reflect surrounding neighborhood properties. Second, the continuity of local asymmetries. It is important to note that within a functional region the continuity of the local vector field is defined and deterministic, as a local slide-vector is uniformly imposed to the local field as a constraint. However, the continuity of the vector field is undefined between the functional regions, as no external constraints are imposed to the asymmetric relationship between the positions of locations in different functional regions in addition to their functional distances. Consequently, the continuity between functional regions in the vector field must be assumed to guarantee the interpolation process is grounded.

To inference potential field from vector field, the observed vector field must satisfy the condition that it can be decomposed into divergence-free and curl-free components. Taking an approach same as Tobler (1975), by calculating the divergence of the vector field, a potential field $a(x, y)$ can be determined through Poisson's equation as Equation (9) (The detailed solution process is provided in Appendix 1).

$$\nabla^2 a = \frac{\partial c}{\partial x} + \frac{\partial c}{\partial y}. \tag{9}$$

Based on the above methods, the derived potential field resembles a landscape of movement potentials, where high elevation of the terrain has a high potential. It is the potential difference that acts as the driving force behind the movements from high potential to low potential locations.

## 4  Data and experiments

### 4.1 Data and study area

We apply the proposed method to analyze the interstate migration flows among the 48 contiguous states of the United States (excluding Alaska and Hawaii). The migration flow data used in this study is obtained from the United States Census reports for four

time periods: 1965-1970, 1975-1980, 1985-1990, and 1995-2000. Based on the flow data that represent the similarity between locations, we apply the reverse estimation of the doubly-constrained spatial interaction model (Plane 1984) to obtain asymmetrical functional distances.

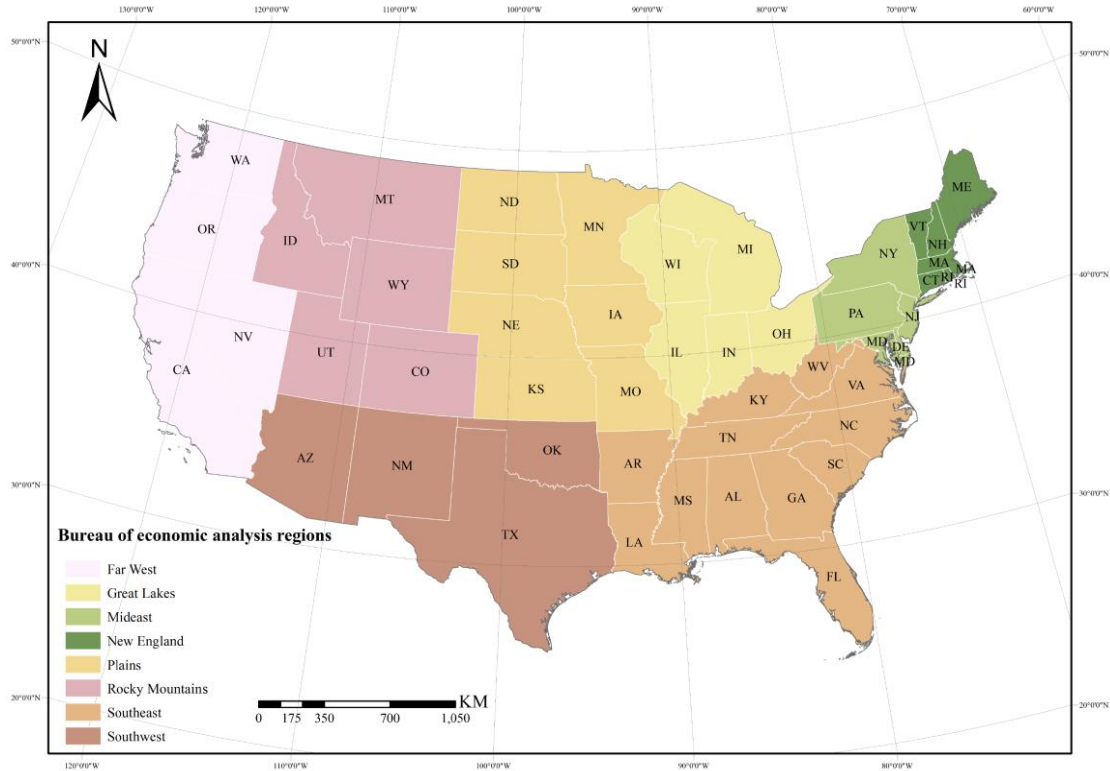

Figure 4. Study area, states and Bureau of Economic Analysis (BEA) regions.

*4.2 Stability evaluation of model results and temporal comparison*

As a special multi-dimensional unfolding model, SVM is susceptible to solution instability (Borg and Groenen 2005). As the model mainly relies on data for a subset of dyads (those between O and D entities, none among O or D entities) as constraints, it is susceptible to indeterminacy (Borg et al. 2018). Moreover, this instability is scale-dependent: as the SVM model shifts from individual to global levels and degrees of freedom decrease, the instability issue of models may ease out.

To evaluate the stability of the generalized SVMs for delineating the asymmetry in space and for comparing the asymmetry over time, we used a series of metrics and designed a decision procedure from multiple runs with randomized initial conditions to evaluate the local SVMs' results. We considered three aspects to identify stable results: 1) badness-of-fit; 2) stability across multiple runs with randomized initial conditions; and 3) the similarity of the embedding configurations with geographical configurations when the first two criteria are met. Several evaluation metrics were introduced as follows:

(1) Badness-of-fit: The fitted slide-vector models are assessed by metrics based on the stress value, which minimizes the discrepancy between the model-derived distances $d_{ij}(X,Y)$ and the observed asymmetry spatial separations $\delta_{ij}$ (Sturrock and Rocha 2000). We adopt a simplified definition of two common badness-of-fit statistics: Normalized Raw Stress (NRS) and the squared form of Kruskal's STRESS formula one ($Stress_I^2$).

$$NRS = \frac{\sum \left(\delta_{ij} - d_{ij}(X,Y)\right)^2}{\sum \delta_{ij}^2} \quad (10)$$

$$Stress_I^2 = \frac{\sum \left(\delta_{ij} - d_{ij}(X,Y)\right)^2}{\sum d_{ij}(X,Y)^2} \quad (11)$$

(2) Bidimensional correlation: Introduced by Tobler (1994), this statistic extends traditional regression to two dimensions, analyzing the correlation between two spaces represented by two dimensional configurations.

We established evaluation procedures for different versions of SVMs. For the individual-level SVM, which has the highest degree of freedom and influences the constraints of the local SVM, we conducted 50 randomized experiments, each involving 1,000 random initializations to select the optimal result. For the local and global SVMs, a simplified evaluation procedure, analogous to that of the individual-level SVM, was designed (Due to space limitations, please refer to Appendix 2 for details).

To ensure the comparability of embedding results across time, two steps were undertaken. First, the functional distances $\delta_{ij}$ used in the SVM model were linearly transformed to a range of 0 to 5 while maintaining proportionality with the original distances. Second, Procrustes analysis (Gower 1975; Wang and Mahadevan 2008) was applied to align the SVM-derived configurations with the two-dimensional configurations obtained from symmetrical geographical distances (Appendix 3). Here Procrustes analysis can be utilized as a normalization process for two-dimensional embeddings. It involves translation, rotation, and scaling to align the SVM's configurations into correspondence with the geographical configurations used as a reference, without altering the spatial relationships between slide-vectors.

## 5  Results and discussions

### *5.1 Asymmetry of migration space at different levels*

To answer Q1, we examine the extent to which the local SVMs can effectively capture the spatial dependency and heterogeneity of asymmetries. We use the functional distances derived from migration data of 1995-2000 as a case study.

We start by looking at model performance for the two-dimensional solutions of the global, local and individual-level SVMs using badness-of-fit measures defined by Equation (10-11). As shown in Table 1, the local SVM outperforms the global version, as it captures the heterogeneity of asymmetry at the local level. Understandably, the individual-level SVMs achieve the lowest badness-of-fit by imposing individual-level constraints. However, they focus excessively on the specificity of distance structures associated with each individual state and easily overfit the data. Therefore, they cannot capture the common structures of asymmetries at larger scales.

Table 1. Overall badness-of-fit

|  | Global | Local | Individual |
|---|---|---|---|
| Normalized Raw Stress (NRS) | 0.07928 | <u>0.07793</u> | **0.07745** |
| $Stress_I^2$ | 0.08611 | <u>0.08431</u> | **0.07943** |

*Note: Bold components are the best model, underlined are the second best.

Following Tobler (1975), we split the distance matrix into symmetric and asymmetric parts. Given a distance matrix $P$ and its transpose $P^T$, the symmetric part is defined as $\frac{P+P^T}{2}$, and the asymmetric part as $\frac{P-P^T}{2}$. To quantify the badness-of-fit contributed by each component, we adopt the NRS metric as it uses the original dissimilarity matrix as the denominator, making it more suitable for decomposition and model comparison. The stress is decomposed as:

$$NRS_{total} = \frac{RSS_{sym} + RSS_{asy}}{DSS_{total}} = w_{sym} \cdot NRS_{sym} + w_{asy} \cdot NRS_{asy} \quad (12)$$

where $RSS_{asy}$ and $RSS_{sym}$ denote the residual sum of squares (stress) for the symmetric and asymmetric components, respectively, and $DSS_{total} = \sum \delta_{ij}^2$ is the dissimilarity sum of squares of the transformed distance matrix. The weights $w_{sym}$ and $w_{asy}$ are the proportions of $DSS_{sym}$ and $DSS_{asy}$ to $DSS_{total}$, respectively. Detailed derivations are provided in Appendix 4.

The relative contribution of each part is then defined as：

$$Contribution_{sym} = \frac{w_{sym} \cdot NRS_{sym}}{NRS_{total}}, Contribution_{asy} = \frac{w_{asy} \cdot NRS_{asy}}{NRS_{total}} \quad (13)$$

The relative contribution indicates the proportion of total errors attributable to each component. A lower contribution for a given component implies that less errors are associated with that component and the model fitted that component of the data more effectively.

Table 2. Contribution to badness-of-fit of the symmetrical and asymmetrical parts

| Component | Weight | Contribution | | |
|---|---|---|---|---|
| | | Global | Local | Individual |
| Symmetry | 0.9778 | **0.72026** | 0.73534 | <u>0.72285</u> |
| Asymmetry | 0.0222 | 0.27974 | **0.26466** | <u>0.27715</u> |

*Note: Bold components are the best model, underlined are the second best.

As shown in Table 2, the local SVM contributes the least to the overall asymmetric error, reflecting its best fit for the asymmetric structure. By imposing local constraints, the local SVM avoids overfitting by balancing both components and captures the heterogeneity of local asymmetries. In contrast, the global model, while fitting best on the symmetric part, shows poor performance in capturing asymmetric structures, when the data exhibits local asymmetry rather than a global one. The individual-level SVM performs moderately in fitting both the symmetric and asymmetric components. This is because it naturally prioritizes the dominant symmetric structure in its fine-grained fitting, while, due to its individually defined specification for the heterogeneity of asymmetries, it also focuses on the fitting of the asymmetric part.

Figure 5(a, b, and c) presents the fitted global, local, and individual-level SVMs, respectively, revealing the corresponding asymmetric patterns of migration space at different levels. The embedded positions of each state are represented by a vector as an arrow, with the head and tail indicating the origin and destination positions respectively. The length of the vector indicates the relative magnitude of asymmetry. In Figure 5 (b) and (c), the local regions are functional regions based on the grouping of individual slide vectors showing similar asymmetric patterns (the determination procedure was specified in Section 3.2.2), which are indicated by the color-coded vectors.

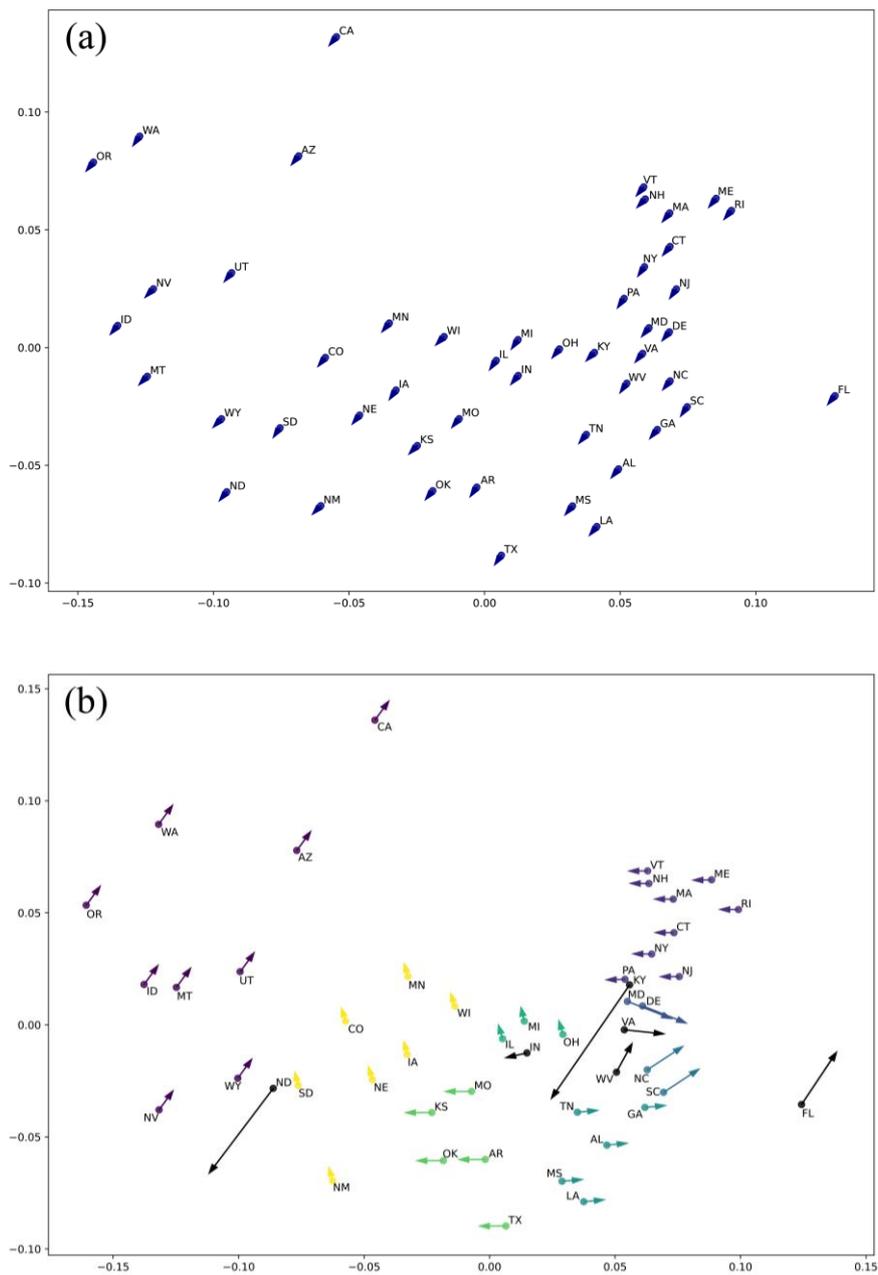

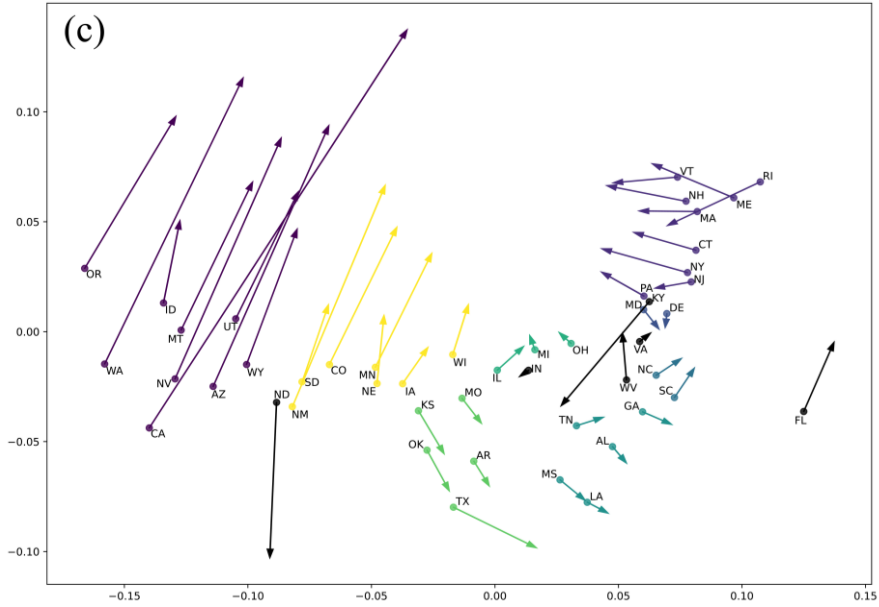

Figure 5. Comparison of three slide-vector models: (a) global, (b) local, and (c) individual-level SVMs.

Table 3. Local versus global slide-vectors.

| Rank | Vector length* | Model | |
|---|---|---|---|
| | | States of local SVM | States of global SVM |
| 1 | 0.05551 | KY | |
| 2 | 0.04174 | ND | |
| 3 | 0.02329 | FL | |
| 4 | 0.01512 | DE、MD | |
| 5 | 0.01363 | NC、SC | |
| 6 | 0.01199 | VA | |
| 7 | 0.01030 | WV | |
| 8 | 0.00657 | MO、KS、OK、TX、AR | |
| 9 | 0.00625 | WA、MT、ID、OR、WY、CA、NV、UT、AZ | |
| 10 | 0.00479 | IN | |
| 11 | 0.00418 | TN、AL、GA、MS、LA | |
| 12 | 0.00347 | ME、NH、VT、NY、MA、PA、CT、RI、NJ | |
| 13 | 0.00233 | MI、IL、OH、MN、WI、SD、IA、NE、CO、NM | |
| 14 | 0.00076 | -- | Global |

*Note: All vector lengths have been uniformly scaled for comparison.

The local structures of asymmetry are demonstrated in Figure 5(b), which sharply contrasts to the weak global trend of asymmetry in both magnitude and direction in Figure 5(a). To gain deeper insights into the spatial heterogeneity of local asymmetry, we compare the magnitude (length) of global and local slide-vectors (Table 3). Results show that the global slide-vector is smaller than all local slide-vectors.

In terms of direction, the local slide-vectors are more pronounced and diverse compared to the global one. As shown in Figure 5(a), the global slide-vector is oriented towards the southwestern states from the northeastern ones. When projecting the origin positions of states onto the global slide-vector represented by a real axis with increasing values towards the arrow side (depicted in Figure 1(c)), it shows that southwestern states play a slightly more important role as origins than as destinations, while the northeastern states do the opposite. In contrast, as shown in Figure 5(b), the directions of the slide-vectors in the local SVM reveal the regional dependencies and heterogeneity of asymmetries that the global model overlooks. For instance, the New England states' vectors mainly point westward, while the vectors of the Western states primarily point towards the Northeast. This asymmetric relationship reflects the regional interaction between the two coastal regions, which serve as each other's migration destinations. Notably, some functional regions correspond to geographic regions, such as the Southern states, the Western states, and the New England states. This correspondence suggests the underlying influence of geographical boundaries on the formation of migration space and the embedded asymmetric patterns.

In Figure 5(c), individual-level SVM results show heterogeneous slide-vectors at the finest level exhibit spatial dependency. Individual slide-vectors with similar directions and magnitudes are clustered by HDBSCAN, implying regional trends of asymmetry. However, as the individual-level model is not designed to capture local dependency, it cannot properly reflect the true regional structures of asymmetry.

Additionally, its individually defined specification for the heterogeneity of asymmetries leads to a high degree of freedom, which may cause instable results. Therefore, it is imperative to have a properly defined local SVM that is tailored to spatial dependency and heterogeneity of asymmetries in the data and produce reliable results.

*5.2 Comparison of migration spaces over time accounting for local asymmetry*

To answer Q2, we utilize stable local SVM embeddings to study how the functional migration space changed over four periods from 1965 to 2000. We first present the temporal characteristics of the local migration space, and then unravel local migration trends by examining the magnitude of asymmetry (Section 5.2.1) and the spatial topological patterns of distinct asymmetric O position and D position embedding spaces (Section 5.2.2).

Figure 6 (a, b, c, and d) displays the local SVM embeddings for four time periods. All local SVM results have passed the stability tests described in Section 4.2. Regions of local asymmetry are differentiated by color and identified by number. Black vectors represent states that do not belong to any region in the local model and are individually constrained instead, each one possessing its own distinct slide-vector.

As depicted in Figure 6, the local asymmetry of migration space exhibits a certain stability over time (The results for corresponding individual-level SVMs are included in Appendix 5). From the geographical perspective, the asymmetric local regions broadly correspond with the geographical topology. Specifically, the regions embedded on the right-hand side correspond to the New England, Mideast, and Southeast regions of the east coast of the United States. On the left-hand side, we find the Far West, Rocky

Mountains, and Southwest regions of the west coast. The center of the embedded space is distributed the Great Lakes and Plains regions. Additionally, at the edges of these geographical divisions, there is evident temporal convergence and restructuring by states among the asymmetrical regions.

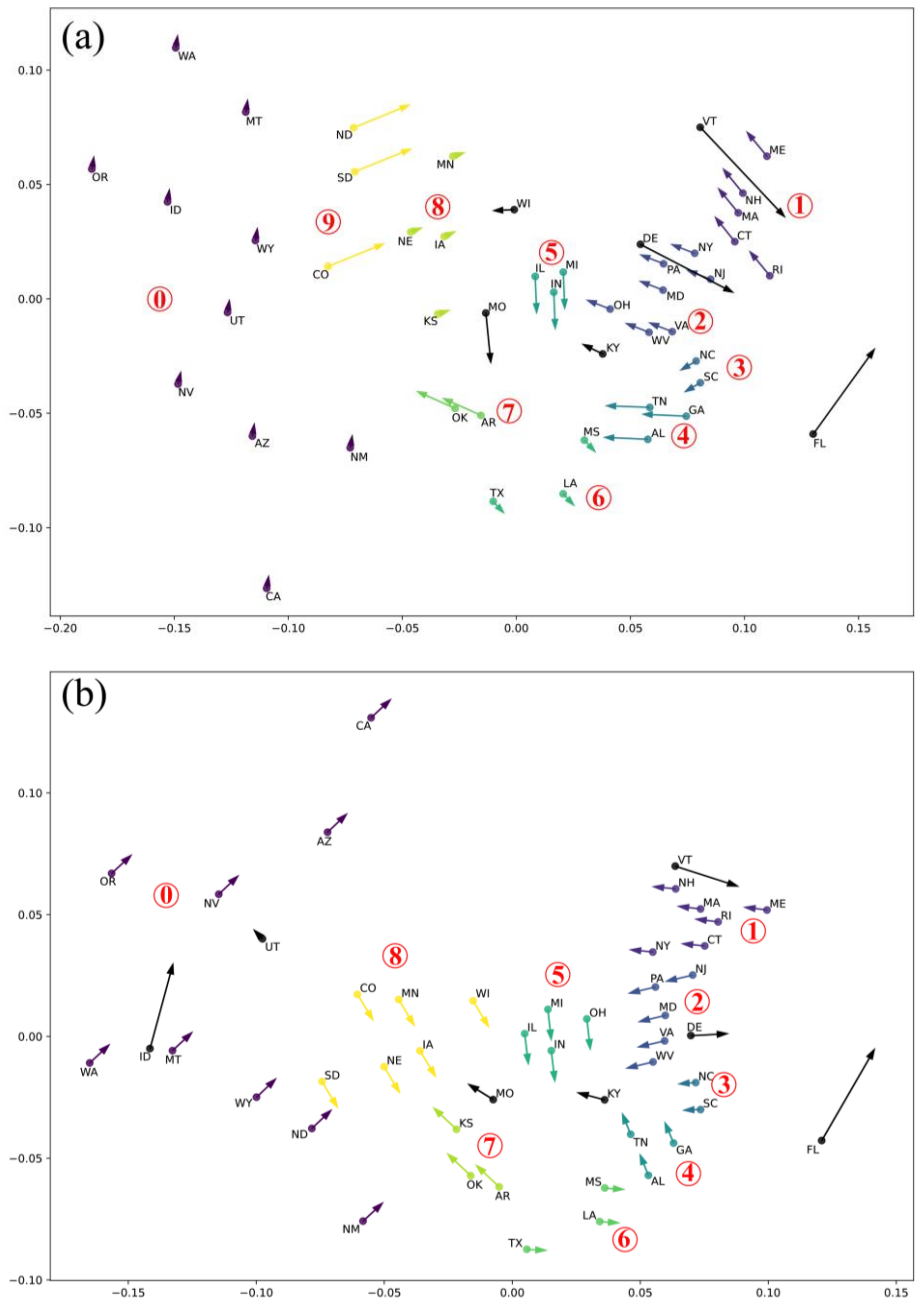

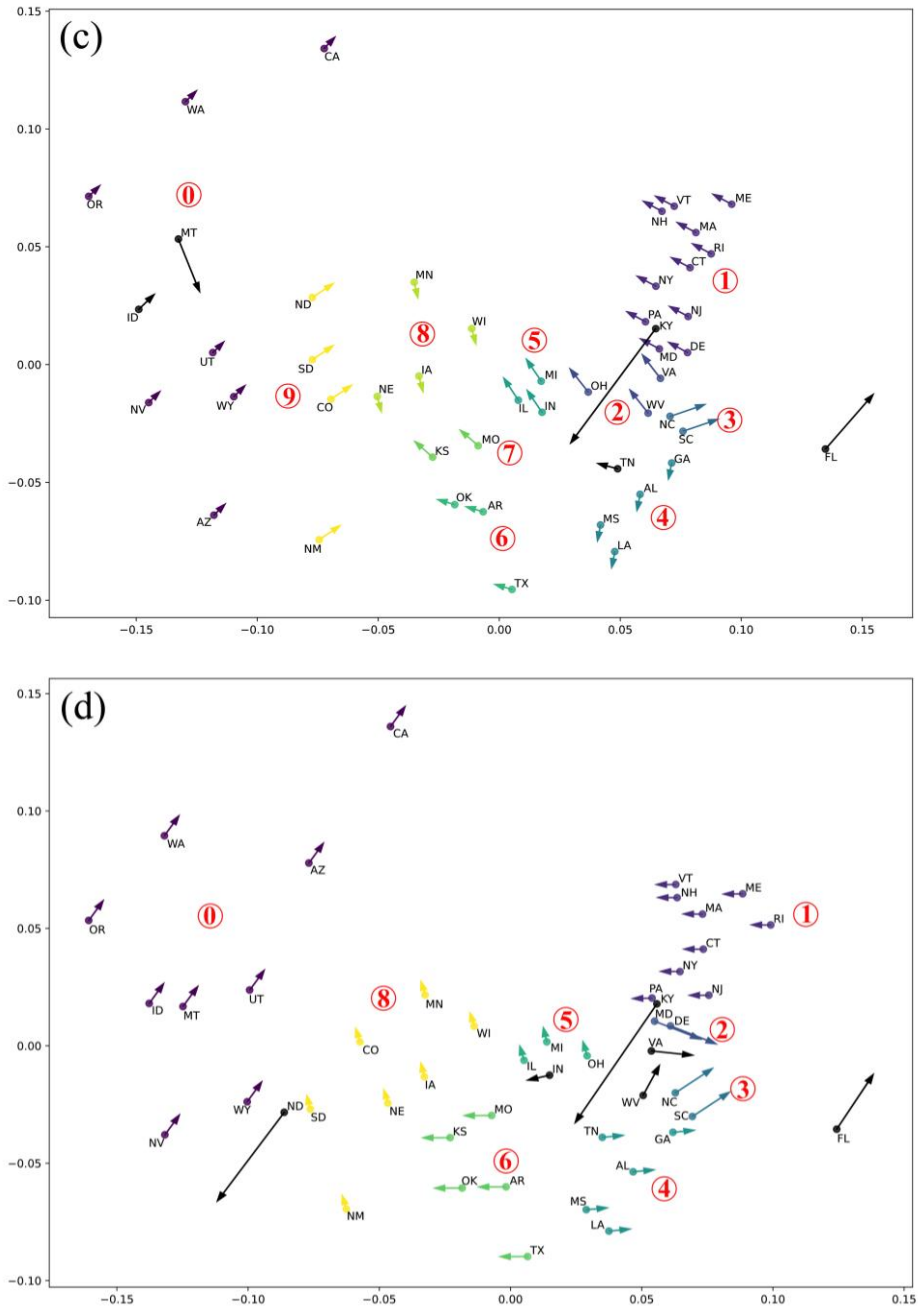

Figure 6. Local slide-vectors in time series from 1965 to 2000: (a) 1965-1970, (b) 1975-1980, (c) 1985-1990, and (d) 1995-2000. Regions of local asymmetry are differentiated by color and identified by number.

*Magnitude of local asymmetry*

To clearly illustrate changes in the relative asymmetries between local regions, we adopt the length of the slide-vector as a metric for the magnitude of asymmetry.

Hypothetically, if a state's origin and destination positions were to overlap perfectly, it would imply a symmetrical migration pattern where the migration resistance (functional distance) for migrations towards and from that state is identical. Ideally, this symmetry would manifest in equal migration inflows and outflows. On the other hand, deviations from this symmetry -- the asymmetries indicated by a longer slide-vector -- theoretically correspond to larger net migration flows in reality.

Figure 7 employs a Sankey diagram to visualize these dynamics. Each rectangular block on the vertical axes represents a local slide-vector region of states or an individual state. Nodes on each vertical axis (for a specific period) are ranked in descending order of the length of slide-vectors from top to bottom, thus facilitating the comparison of the magnitude of local asymmetries across time. Nodes comprising roughly the same states are consistently color-coded across all four time periods. Figure 8 demonstrates that regions with high asymmetry generally remained in high asymmetry over the four time periods, although there were changes of local states in those regions; the same happens to regions with low asymmetry. Some regions experienced fluctuations in overall local asymmetry (such as cluster 0: Far West and Southwest, cluster 1: New England, and cluster 5: Great Lakes), indicating changes in the contrast between their migration inflows and outflows over time.

Notably, the series of Sankey diagrams clearly shows the reorganization of local asymmetry regions. Notably, while most clusters remain stable, significant changes occur near their margins. For instance, cluster 8, situated at the intersection of the Plains and Great Lakes regions, exhibits multiple swaps with cluster 9 (Plains, Rocky

Mountains and Southwest boundary). This pattern observed at the boundaries implies the existence of geographical border effects of migration asymmetries (Leerkes, Leach, and Bachmeier 2012; Wang et al. 2023). Within the boundaries of distinct economic zones, the local asymmetry regions demonstrate a temporal stability. Nevertheless, at the junctions of heterogeneous areas, these asymmetry migration regions experience corresponding temporal fluctuations.

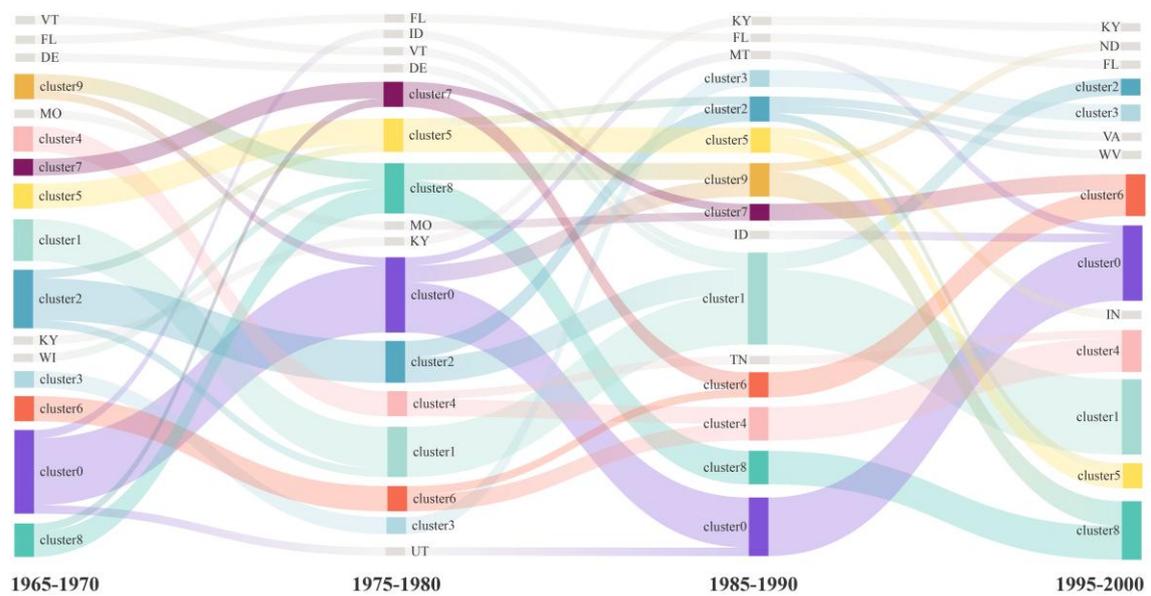

Figure 7. Sankey diagram of the length of slide-vectors from local SVMs over time.

*Configurations of local asymmetry in the migration space*

Here we focus on the functional configurations of states (via embeddings), as migration origins and destinations, respectively, by examining their association via local asymmetries in the migration spaces and the structural changes over time. First, from the configurations of the local SVM model, we extract the asymmetric migration spaces composed of migration origin configurations and destination configurations, referred to as the Origin Position Migration Space (OPMS) and the Destination Position

Migration Space (DPMS), respectively. Second, using the bidimensional regression correlation metric, we measure the similarity between these two migration spaces and the geographical space (Constructing the geographical space configuration detailed in Appendix 3).

Table 4. Bidimensional correlations between asymmetric migration space and geographical space

| Year | Bidimensional correlations | |
|---|---|---|
| | OPMS | DPMS |
| 1965-70 | **0.91787** | **0.90162** |
| 1975-80 | 0.71379 | 0.68532 |
| 1985-90 | <u>0.82810</u> | <u>0.81401</u> |
| 1995-00 | 0.73764 | 0.70484 |

Figure 8 illustrates the OMPS (panels a1, b1, c1, d1) and DPMS (panels a2, b2, c2, d2) over the period of 1965-2000. In both sets of embedded spaces, proximity serves as an indication of migration similarity. Therefore, in the context of migration decisions, neighboring origin states within the OPMS may be subject to similar economic, social, or environmental factors that drive people to migrate from these states. These factors could include a decrease in job opportunities, an increase in living costs, or a decline in environmental quality. Similarly, spatial proximity within the DPMS indicates similarities in destination choices during migration decisions.

The comparison of the spatial configuration between the OPMS and the DPMS in Figure 8 reveals several key insights. First, both Origin and Destination spaces exhibit spatial proximity relationships corresponding to geographical space, thus emphasizing the role of geographical distances in migration. However, they also reveal proximity patterns that deviate from those of symmetrical geographical space. Specifically, Table 4 shows that the migration space of 1965-1970 (Figure 8a) closely resembles the original geographical topology (BDR = 0.91). However, by 1975-1980 (Figure 9b), notable distortions emerged, particularly near the West Coast, with states like California, Nevada and Utah bending upwards towards the East Coast. The Southwest region fragments, with Arizona appearing closer to California and Nevada, while Oklahoma, Texas and New Mexico retain adjacency with the Southeast. This distortion leads to a decreased correlation with geographical space (0.71 and 0.68). During 1985-1990 (Figure 8c), trends do not intensify, however; the Southwest returns to its original pattern, and the Rocky Mountains and Plains show increased proximity and a nested pattern. By 1995-2000 (Figure 8d), the overall inter-regional adjacency becomes more compact, further distorting the spatial topology, akin to a folding and twisting of the US map along the north-south axis.

Second, the results of the BDR in Table 4 demonstrate that the OPMS exhibits higher similarity to geographical space than DPMS in different time periods. This may reflect that the factors influencing decision to migrate may be more spatially dependent, while migration destination choice may be influenced by broader socio-economic factors, that are less geographically related. Non-geospatial factors, such as career development, and family circumstances, which are more closely related to overall living environment and changing individual/family needs (Rossi and Shlay 1982).

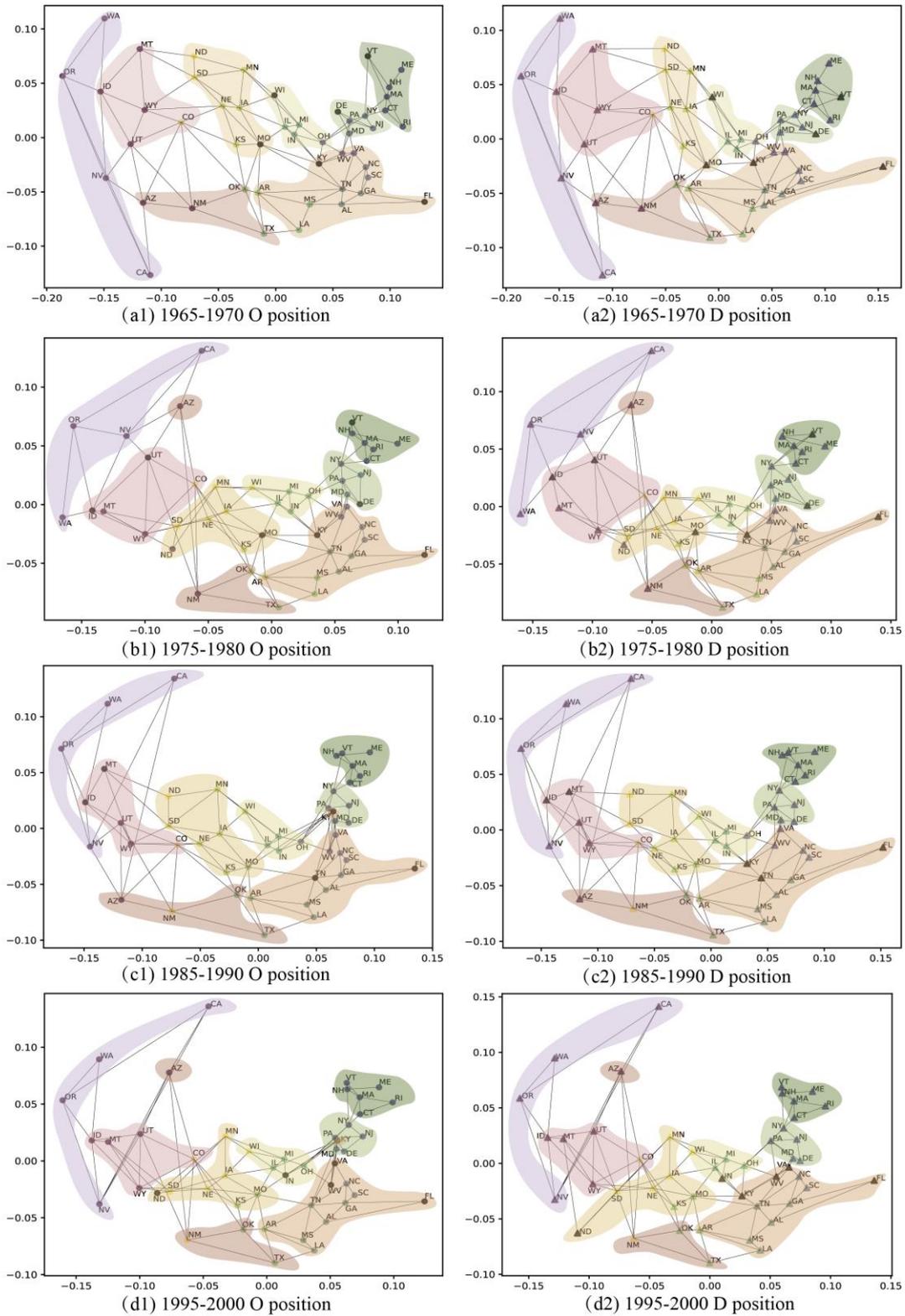

Figure 8. Proximity patterns considering the spatial asymmetry. The black lines represent queen neighbors of the original U.S. state boundaries indicating the geographical adjacency relationships.

## 5.3 Migration-scape: constructing the latent migration resistance landscape

To answer Q3, we first transform the migration space accounting for local asymmetries to a continuous landscape representation of migration resistance, referred to as a migration-scape. This transformation involves performing vector field interpolation on local slide-vectors and deriving a potential field, as detailed in Section 3.3. Then, we assess whether the migration potential field can be used to infer asymmetric migration patterns between local regions by comparing the inferred migration resistance (inter-states potential difference) with the original migration resistance (functional distance).

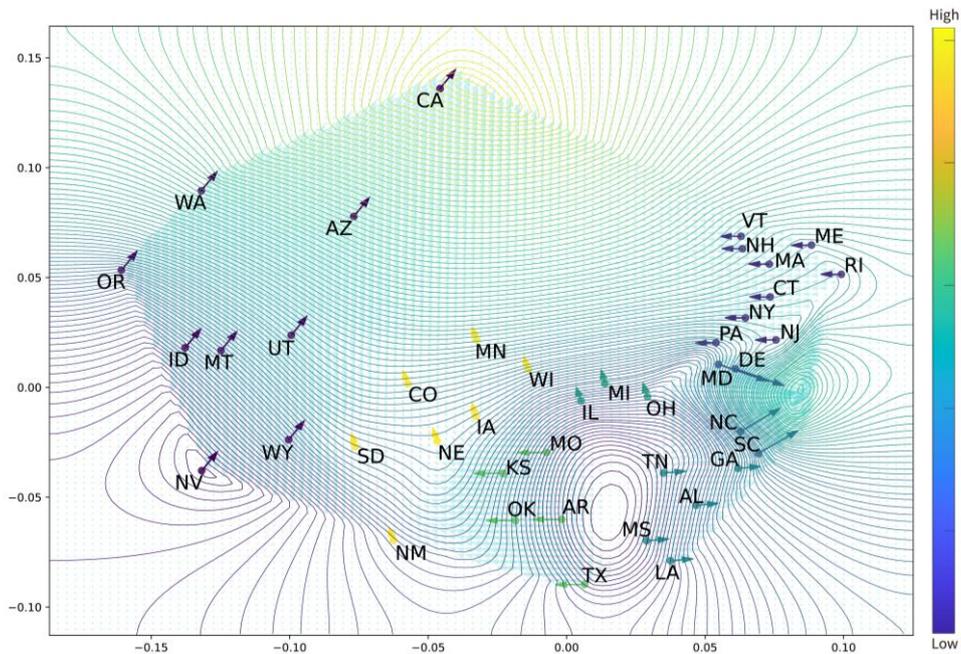

Figure 9. Vector and potential fields in the 1995-2000 migration space.

Figure 9 shows both the vector field and the migration-scape produced from local slide-vectors based on the 1995-2000 data. The migration-scape is depicted by equipotential lines with a color gradient from blue to yellow, indicating low to high potential values. Each colored arrow symbolizes a local slide-vector associated with the positions of a state; they serve as the seed vectors for vector interpolation and the un-clustered individual slide-vectors (black vectors in Figure 6 (d)) are not included in the interpolation. The interpolated vector field is depicted by cyan vectors.

Figure 10 shows the destination positions within the migration space from selected viewpoints. Here, the Z-axis represents the potential values. The X and Y axes show the coordinates of the destination positions. We chose to assign the potential values to the destination positions because each destination's potential takes into account the length of a slide-vector, capturing the magnitude of the asymmetry. These potential values are analogous to elevations in a mountainous terrain, where traveling to high potential (yellow color) areas of the migration-scape is more costly than to the low potential (blue color) areas.

The instances of migration-scape for three periods of 1965 to 1980 show great heterogeneity in the east-west direction. States in the Great Lakes and Plains regions are situated in areas of high potential, while states located in east and west coast regions are in areas of low potential (Figure 10a-c). This observation suggests a ridge of high potential in the central region, dividing the eastern and western areas, and reflects a general migration trend from the central towards two coastal regions. Compared to this, the instance of migration-scape for 1995-2000 exhibits less heterogeneity in the east-west direction, which shows a trend of potential increasing from high to low in the north-south direction. This suggests that during this period, migration flowed more prominently from northern states to southern states, while the trend of migration from the central regions towards the two coastal regions diminished.

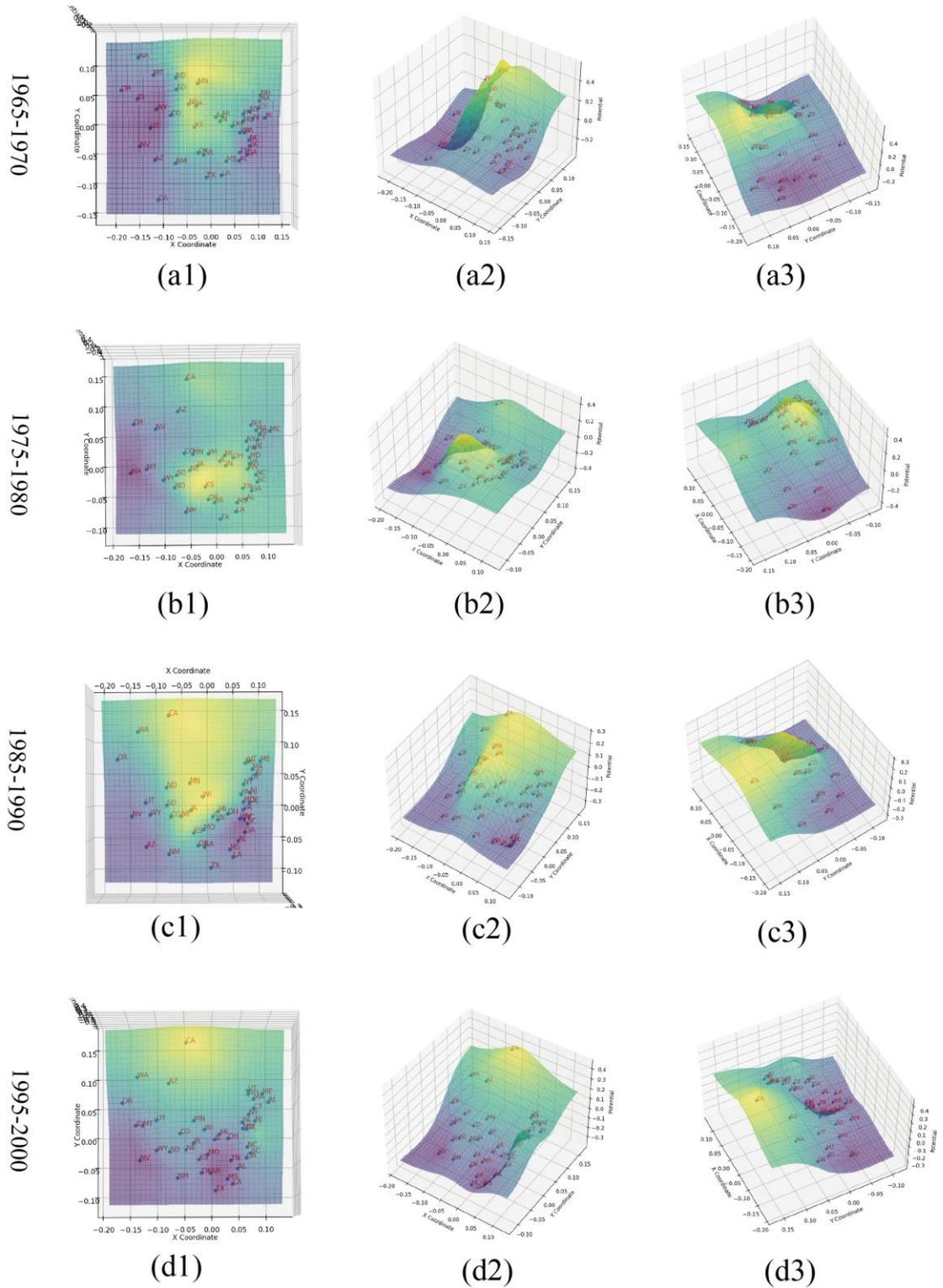

Figure 10. Potential of migration-scape from 1965-2000. (1), (2) and (3) represent three different viewpoints of the migration-scape. The blue dots indicate the destination positions of states.

Figure 11 shows the temporal evolution of the heterogeneity of local asymmetries across four time periods. We normalized the potential values at destination positions, then categorized them into four groups by K-Means (Kodinariya and Makwana 2013). The four groups can be identified from the evolution of local asymmetries, which broadly correspond to the geographical regions of the Southwest to Southeast Mixed Region, Northeast, Midwest, and South. Specifically, the Midwest Region maintained a high potential during 1965-1990 but experienced a decline in potential during 1995-2000. This trend aligns with the population outflow from the Rust Belt and the recovery policies initiated in the late 1990s. Furthermore, it corresponds to the pattern presented in Figure 10. The Southwest to Southeast Mixed Region and the South Region demonstrate high potential in 1965-1980, followed by a decrease in the next two periods. On the other hand, the Northeast and South Regions maintain a relatively moderate potential throughout.

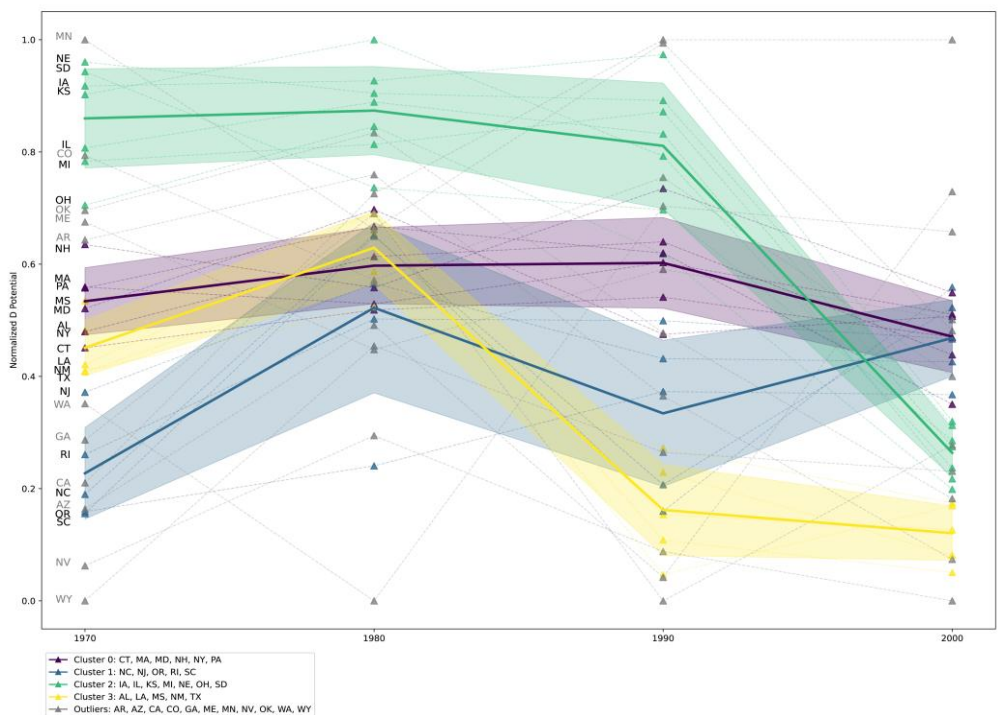

Figure 11. Local asymmetry potential of migration-scape in time series from 1965 to 2000.

We then evaluated whether the migration-scape can be used to effectively infer the inter-regional asymmetry patterns, which are unconstrained by the local SVMs. In the migration-scape, we postulated that there exist 'mountain ridges' representing barriers that prevent flows from separated 'watersheds' of migration. Based on this analogy, we hypothesized that the migration-scape is more continuous within a 'watershed' than between two 'watersheds'; hence, the migration-scape should do a better job in inferring migration resistance between different local asymmetric regions within a 'watershed' than doing the same between 'watersheds'.

To test this hypothesis, we identified typical 'watersheds' in the instances of migration-scape (Appendix 6). To evaluate the goodness of asymmetry inference between local asymmetric regions based on the migration-scape, we devised a directional consistency index (DCI), which compares the inferred net migration resistance (difference in potential between two locations) to the real net migration resistance (difference between functional distances in opposite directions). DCI measures the consistency (1 as true and 0 as false) between the sign of the two net resistances between locations in the migration-scape. When the real net migration resistance from A to B is negative, meaning that traveling from A to B has lower resistance than traveling in the opposite direction. Consequently, the potential difference from A to B must be positive (A is higher than B in the migration-scape). Technically, we first selected two local asymmetry regions (either in one 'watershed' or in different 'watersheds'; second, we calculated the DCI between the two net migration resistance for all combinations of pair locations; third, we calculated the proportion of matching pairs, whose DCI value is 1, as an aggregate indicator (DCR) for the goodness of asymmetry inference between the two regions (details of DCI and DRC are described in Appendix 7).

Table 5 shows that the average DCR within a single 'watershed' is consistently higher than that between two 'watersheds'. This is valid across all time periods except for 1995-2000 where a significant watershed was not observed. The results support our hypothesis that the migration-scape is more effective in inferring migration resistance within a 'watershed', indicating a stronger consistency and accuracy in capturing migration patterns within a single watershed.

Table 5. Averaged DCR for different local asymmetric regions within one 'watershed' and between two 'watersheds', respectively.

| Year | Global | Local | |
|---|---|---|---|
| | | Intra-watershed | Inter-watersheds |
| 1965-70 | 0.5091 | 0.5445 | 0.4824 |
| 1975-80 | 0.5277 | 0.5778 | 0.4923 |
| 1985-90 | 0.5349 | 0.5656 | 0.4857 |
| 1995-00 | 0.5306 | -- | -- |

*Note: The migration-scape from 1995 to 2000 does not exhibit a significant watershed.

The significance of this hypothesis lies in its implication for understanding underlying migration barriers. Better inference achieved within 'watersheds' than between them suggests that the migration patterns are relatively more similar within these regions. This could be due to shared socio-economic, cultural, or environmental factors that influence migration decisions within a watershed. On the other hand, the presence of ridges or barriers in migration-scape indicates a drastic change in migration pattens akin to a discontinuity in the migration-scape, possibly caused by resources, opportunities, or cultural differences between the two sides of a barrier. Therefore, the migration patterns between 'watersheds' may exhibit greater variability and complexity compared to what prevails within a single 'watershed'. As illustrated in Figure 10, migration patterns during 1965-1970, 1975-1980 and 1985-1990 exhibited a well-defined barrier between the eastern and western regions, while this barrier eased to be less pronounced in 1995-2000. Identifying these 'barriers' could lead to valuable insights into the underlying factors driving migration.

## 6 Conclusions and future work

Current methods of spatial analysis struggle to effectively model the structural asymmetry inherent to functional space. This study undertook in-depth examination on local asymmetry in spatial interactions and has made significant contributions in three main aspects. First, a novel approach was proposed for tackling the challenge of modeling asymmetry of spatial interactions. Specifically, we introduced a local slide-vector model that incorporates the spatial dependency and heterogeneity of asymmetry in functional spaces. The model effectively captures local asymmetric structures of spatial separations through the utilization of spatially constrained multi-dimensional unfolding. Experimental and visual validation provided evidence on the effectiveness of the proposed model in preserving asymmetry at local.

Secondly, we introduced a potential field approach to representing and visualizing the asymmetric structures of functional spaces. By transforming the discrete embedding space into a continuous potential field, referred to as 'migration-scape' in the case study, this approach revealed meaningful patterns of migration resistance. We evaluated that to what extent the migration-scape can be used to effectively infer the inter-regional asymmetry patterns of migration by identifying barriers akin to the discontinuity in the migration-scape.

Thirdly, in the case study of the U.S. migration space, our findings revealed that functional migration separations are often asymmetric and very different from absolute physical distance. The asymmetry of migration space influenced by non-geographical factors, exhibits functional distortions that deviate from geographical space. However, notably, the local asymmetry structures exhibit temporal stability while reflecting regional reorganizations, which closely align with geographical boundaries. It highlights that the geographical space, particularly the constraints imposed by physical distances, remains a significant factor shaping the landscape of U.S. migration space.

Future studies involve exploring interactive tools or machine learning algorithms to extract and highlight meaningful patterns from the asymmetry data. Moreover, investigating the relationship between asymmetric distances and other geographical factors, such as socioeconomic status or transportation and communication networks, will help to yield deeper insights into the underlying mechanisms governing spatial interactions.

## References

Ahmed, Nobbir, and Harvey J. Miller. 2007. Time–Space Transformations of Geographic Space for Exploring, Analyzing and Visualizing Transportation Systems. *Journal of Transport Geography* 15 (1):2–17.

Allard, Antoine, M. Ángeles Serrano, and Marián Boguñá. 2024. Geometric Description of Clustering in Directed Networks. *Nature Physics* 20 (1). Nature Publishing Group:150–156.

Ben-Elia, Eran, Roberta Di Pace, Gennaro N. Bifulco, and Yoram Shiftan. 2013. The Impact of Travel Information's Accuracy on Route-Choice. *Transportation Research Part C: Emerging Technologies* 26 (January):146–159.

Bongiorno, C., Zhou, Y., Kryven, M., Theurel, D., Rizzo, A., Santi, P., Tenenbaum, J., Ratti, C., 2021. Vector-based pedestrian navigation in cities. *Nature Computational Science* 1, 678–

685.

Borg, Ingwer, and Patrick JF Groenen. 2005. *Modern Multidimensional Scaling: Theory and Applications*. Springer Science & Business Media.

Borg, Ingwer, Patrick JF Groenen, and Patrick Mair. 2018. Applied Multidimensional Scaling and Unfolding. Springer.

Broitman, Dani, Karima Kourtit, Peter Nijkamp, and Waldemar Ratajczak. 2021. Gravitational Analysis in Regional Science and Spatial Economics: A Vector Gradient Approach to Trade. *International Regional Science Review* 44 (3–4). SAGE Publications Inc:400–431.

Bunge, William. 1966. *Theoretical Geography*. Lund : Royal University of Lund, Dept. of Geography ; Gleerup.

Chino, Naohito. 1978. A Graphical Technique for Representing the Asymmetric Relationships between N Objects. *Behaviormetrika* 5. Springer:23–40.

Christakos, George. 2000. *Modern Spatiotemporal Geostatistics*. Vol. 6. Oxford university press.

Cliff, Andrew D., and Peter Haggett. 1998. On Complex Geographic Space: Computing Frameworks for Spatial Diffusion Processes. In *Geocomputation: A Primer*, 231–256. John Wiley and Sons, New York.

Codol, Jean-Paul, Maria Jarymowicz, Marta Kaminska-Feldman, and Anna Szuster-Zbrojewicz. 1989. Asymmetry in the Estimation of Interpersonal Distance and Identity Affirmation. *European Journal of Social Psychology* 19 (1). Wiley Online Library:11–22.

Constantine, Alan G., and J. C. Gower. 1982. Models for the Analysis of Interregional Migration. *Environment and Planning A* 14 (4). SAGE Publications Sage UK: London, England:477–497.

Couclelis, Helen. 1999. Space, Time, Geography. *Geographical Information Systems* 1. John Wiley & Sons New York:29–38.

Crivellari, Alessandro, and Euro Beinat. 2019. From Motion Activity to Geo-Embeddings: Generating and Exploring Vector Representations of Locations, Traces and Visitors through Large-Scale Mobility Data. *ISPRS International Journal of Geo-Information* 8 (3). Multidisciplinary Digital Publishing Institute:134.

Damiani, Maria Luisa, Andrea Acquaviva, Fatima Hachem, and Matteo Rossini. 2020. Learning Behavioral Representations of Human Mobility. In *Proceedings of the 28th International Conference on Advances in Geographic Information Systems*, 367–376. Seattle WA USA: ACM.

Doel, Marcus A. 2007. Post-Structuralist Geography: A Guide to Relational Space by Jonathan Murdoch. *Annals of the Association of American Geographers* 97 (4). Routledge:809–810.

Ewing, Gordon. 1974. Multidimensional Scaling and Time-Space Maps. *The Canadian Geographer/Le Géographe Canadien* 18 (2). Wiley:161–167.

Fan, Chao, Yang Yang, and Ali Mostafavi. 2024. Neural Embeddings of Urban Big Data Reveal Spatial Structures in Cities. *Humanities and Social Sciences Communications* 11 (1). Palgrave:1–15.

Goodchild, Michael F., May Yuan, and Thomas J. Cova. 2007. Towards a General Theory of Geographic Representation in GIS. *International Journal of Geographical Information Science* 21 (3). Taylor & Francis:239–260.

Gower, John C. 1966. Some Distance Properties of Latent Root and Vector Methods Used in Multivariate Analysis. Biometrika 53 (3–4). Oxford University Press:325–338.

Gower, J. C. 1975. Generalized Procrustes Analysis. Psychometrika 40 (1):33–51.

Getis, Arthur, and Janet Franklin. 2010. Second-Order Neighborhood Analysis of Mapped Point Patterns. In Perspectives on Spatial Data Analysis, ed. Luc Anselin and Sergio J. Rey, 93–100.

Ji, Jiahao, Jingyuan Wang, Zhe Jiang, Jiawei Jiang, and Hu Zhang. 2022. STDEN: Towards Physics-Guided Neural Networks for Traffic Flow Prediction. *Proceedings of the AAAI Conference on Artificial Intelligence* 36 (4):4048–4056.

Ji, Jiahao, Jingyuan Wang, Zhe Jiang, Jingtian Ma, and Hu Zhang. 2020. Interpretable Spatiotemporal Deep Learning Model for Traffic Flow Prediction Based on Potential Energy Fields. In *2020 IEEE International Conference on Data Mining (ICDM)*, 1076–1081.

Johnson, Samuel. 2020. Digraphs Are Different: Why Directionality Matters in Complex Systems. *Journal of Physics: Complexity* 1 (1). IOP Publishing:015003.

Jones, Martin. 2009. Phase Space: Geography, Relational Thinking, and Beyond. *Progress in Human Geography* 33 (4). SAGE Publications Sage UK: London, England:487–506.

Kodinariya, Trupti M., and Prashant R. Makwana. 2013. Review on Determining Number of Cluster in K-Means Clustering. International Journal 1 (6):90–95.

Kruskal, Joseph B. 1964. Multidimensional Scaling by Optimizing Goodness of Fit to a Nonmetric Hypothesis. *Psychometrika* 29 (1). Springer:1–27.

Kruskal, Joseph B., and Myron Wish. 1978. Multidimensional Scaling. Sage.

Leerkes, Arjen, Mark Leach, and James Bachmeier. 2012. Borders Behind the Border: An Exploration of State-Level Differences in Migration Control and Their Effects on US Migration Patterns. Journal of Ethnic and Migration Studies 38 (1):111–129.

Li, Xijing, Xinlin Ma, and Bev Wilson. 2021. Beyond Absolute Space: An Exploration of Relative and Relational Space in Shanghai Using Taxi Trajectory Data. *Journal of Transport Geography* 93 (May):103076.

Liu, Yu, Zhengwei Sui, Chaogui Kang, and Yong Gao. 2014. Uncovering Patterns of Inter-Urban Trip and Spatial Interaction from Social Media Check-in Data. PloS One 9 (1):e86026.

Liu, Yu, Shengyin Wang, Xuechen Wang, Yunhao Zheng, Xiaojian Chen, Yang Xu, and Chaogui Kang. 2024. Towards Semantic Enrichment for Spatial Interactions. *Annals of GIS* 30 (2):151–166.

Malpas, Jeff. 2012. Putting Space in Place: Philosophical Topography and Relational Geography. *Environment and Planning D: Society and Space* 30 (2). SAGE Publications Sage UK: London, England:226–242.

Marchand, Bernard. 1973. Deformation of a Transportation Surface. *Annals of the Association of American Geographers* 63 (4). Wiley Online Library:507–521.

MacEachren, Alan M. 1980. Travel Time as the Basis of Cognitive Distance∗. *The Professional Geographer* 32 (1). Routledge:30–36.

Mazúr, Elise, and Johann Urbánek. 1983. Space in Geography. *GeoJournal* 7. Springer:139–143.

Mazzoli, Mattia, Alex Molas, Aleix Bassolas, Maxime Lenormand, Pere Colet, and José J. Ramasco. 2019. Field Theory for Recurrent Mobility. *Nature Communications* 10 (1). Nature Publishing Group:3895.

McInnes, Leland, John Healy, and Steve Astels. 2017. Hdbscan: Hierarchical Density Based Clustering. *J. Open Source Softw.* 2 (11):205.

Midler, Jean-Claude. 1982. Non-Euclidean Geographic Spaces: Mapping Functional Distances. *Geographical Analysis* 14 (3). Wiley Online Library:189–203.

Miller, Harvey J. 2000. Geographic Representation in Spatial Analysis. *Journal of Geographical Systems* 2 (1):55–60.

Miller, Harvey J., and Scott A. Bridwell. 2009. A Field-Based Theory for Time Geography. *Annals of the Association of American Geographers* 99 (1). Routledge:49–75.

Miller, Harvey J. 2004. Tobler's First Law and Spatial Analysis. *Annals of the Association of American Geographers* 94 (2). Routledge:284–289.

Okada, Akinori, and Tadashi Imaizumi. 1987. Nonmetric Multidimensional Scaling of Asymmetric Proximities. *Behaviormetrika* 14 (21). Springer:81–96.

Paasi, Anssi. 2011. Geography, Space and the Re-Emergence of Topological Thinking. *Dialogues in Human Geography* 1 (3). SAGE Publications Sage UK: London, England:299–303.

Papadakis, Emmanuel, Bernd Resch, and Thomas Blaschke. 2020. Composition of Place: Towards a Compositional View of Functional Space. *Cartography and Geographic Information Science* 47 (1). Taylor & Francis:28–45.

Plane, David A. 1984. Migration Space: Doubly Constrained Gravity Model Mapping of Relative Interstate Separation. *Annals of the Association of American Geographers* 74 (2). Taylor & Francis:244–256.

Rossi, Peter H., and Anne B. Shlay. 1982. Residential Mobility and Public Policy Issues: 'Why Families Move' Revisited. *Journal of Social Issues* 38 (3):21–34.

Rothkopf, Ernst Z. 1957. A Measure of Stimulus Similarity and Errors in Some Paired-Associate Learning Tasks. *Journal of Experimental Psychology* 53 (2). American Psychological Association:94.

Selmer, Jan, Randy K. Chiu, and Oded Shenkar. 2007. Cultural Distance Asymmetry in Expatriate Adjustment. *Cross Cultural Management: An International Journal* 14 (2). Emerald Group Publishing Limited:150–160.

Shaw, Shih-Lung, and Daniel Sui. 2020. Understanding the New Human Dynamics in Smart Spaces and Places: Toward a Splatial Framework. *Annals of the American Association of Geographers* 110 (2):339–348.

Shepard, Roger N. 1963. Analysis of Proximities as a Technique for the Study of Information Processing in Man. *Human Factors* 5 (1). SAGE Publications Sage CA: Los Angeles, CA:33–48.

Sheppard, Eric. 2006. David Harvey and Dialectical Space-time. *David Harvey: A Critical Reader*. Blackwell Publishing Ltd Oxford, UK, 121–141.

Sturrock, Kenneth, and Jorge Rocha. 2000. A Multidimensional Scaling Stress Evaluation Table. *Field Methods* 12 (1). SAGE Publications Inc:49–60.

Tao, Ran, and Jean-Claude Thill. 2016. Spatial Cluster Detection in Spatial Flow Data. *Geographical Analysis* 48 (4). Wiley Online Library:355–372.

Thill, Jean-Claude. 2011. Is Spatial Really That Special? A Tale of Spaces. In *Information Fusion and Geographic Information Systems: Towards the Digital Ocean*, ed. Vasily V. Popovich, Christophe Claramunt, Thomas Devogele, Manfred Schrenk, and Kyrill Korolenko, 3–12. Lecture Notes in Geoinformation and Cartography. Berlin, Heidelberg: Springer.

Tobler, Waldo Rudolph. 1961. *Map Transformations of Geographic Space*. PhD diss., University of Washington.

Tobler, Waldo. 1975. Spatial Interaction Patterns. RR-75-019.

Tobler, Waldo R. 1979. Estimation of Attractivities from Interactions. *Environment and Planning A* 11 (2). SAGE Publications Sage UK: London, England:121–127.

Tobler, Waldo. 1993. Three Presentations on Geographical Analysis and Modeling. Citeseer.

Tobler, Waldo Rudolph. 1997. Visualizing the Impact of Transportation on Spatial Relations. In *Western Regional Science Association Meeting, Hawaii*.

Tobler, Waldo. 2000. The Development of Analytical Cartography: A Personal Note. *Cartography and Geographic Information Science* 27 (3). Taylor & Francis:189–194.

Wang, Chang, and Sridhar Mahadevan. 2008. Manifold Alignment Using Procrustes Analysis. In Proceedings of the 25th International Conference on Machine Learning, 1120–1127. ICML'08. New York, NY, USA: Association for Computing Machinery.

Wang, Chenglong, Jianfa Shen, Ye Liu, and Liyue Lin. 2023. Border Effect on Migrants' Settlement Pattern: Evidence from China. *Habitat International* 136 (June):102813.

Wesolowski, Amy, Wendy Prudhomme O'Meara, Nathan Eagle, Andrew J. Tatem, and Caroline O. Buckee. 2015. Evaluating Spatial Interaction Models for Regional Mobility in Sub-Saharan Africa. *PLoS Computational Biology* 11 (7):e1004267.

Yang, Hu, Minglun Li, Bao Guo, Fan Zhang, and Pu Wang. 2023. A Vector Field Approach for Identifying Anomalous Human Mobility. *IET Intelligent Transport Systems* 17 (4):649–666.

Young, Forrest W., Yoshio Takane, and Rostyslaw Lewyckyj. 1978. ALSCAL: A Nonmetric Multidimensional Scaling Program with Several Individual-Differences Options. *Behavior Research Methods & Instrumentation* 10 (3). Springer-Verlag New York:451–453.

Zielman, Berrie, and Willem J. Heiser. 1993. Analysis of Asymmetry by a Slide-Vector. *Psychometrika* 58. Springer:101–114.

Zielman, Bertie, and Willem J. Heiser. 1996. Models for Asymmetric Proximities. *British Journal of Mathematical and Statistical Psychology* 49 (1). Wiley Online Library:127–146.

**Appendix 1: Inference of potential field from vector field**

We first assume the existence of a continuous latent vector field $\vec{c}(x, y)$ from which the fitted local slide-vectors are sampled; then, the latent vector field of local asymmetry can be interpolated. Individual-level slide-vectors that do not fall into any cluster are excluded from the interpolation process, as they solely represent the asymmetry of a single location and do not necessarily reflect surrounding neighborhood properties.

*Interpolation of vectors*

To interpolate discrete vector data into a continuous late vector filed $\vec{c}(x, y)$, the process begins by calculating the vector components for each data point. For a given vector defined by its origin $(x_{ori}, y_{ori})$ and destination $(x_{dest}, y_{dest})$, the horizontal and vertical components are computed as $v_x = x_{dest} - x_{ori}$ and $v_y = y_{dest} - y_{ori}$ The origin and destination points are then combined to form the interpolation dataset, where the corresponding vector components are duplicated to match the combined points. Next, a uniform grid is defined over the spatial range of the data, with $x$ and $y$ coordinates spanning from their minimum to maximum values. Using linear interpolation, the vector field components are computed on this grid as $v_x^{interp}(x, y)$ and $v_y^{interp}(x, y)$. Finally, the continuous latent vector field is constructed by combining these interpolated components, resulting in $\vec{c}(x, y) = (v_x^{interp}(x, y), v_y^{interp}(x, y))$.

*Inferring potential field from vector filed*

Following the approach of Tobler (1975), a potential field $a(x, y)$ can be derived from a vector field $\vec{c}(x, y)$, by computing its divergence and solving Poisson's equation $\nabla^2 a = \nabla \vec{c}$. Specifically, the divergence of the vector field $\nabla \vec{c}$ is calculated numerically by Equation (1):

$$\nabla \vec{c} = \nabla^2 a = \frac{\partial c}{\partial x} + \frac{\partial c}{\partial y} \tag{1}$$

where gradients $\frac{\partial c}{\partial x}$ and $\frac{\partial c}{\partial y}$ are approximated using central difference schemes. This method calculates the gradient at each grid point by taking the difference between neighboring points on either side, providing second-order accuracy in the interior of the grid and using one-sided differences at the boundaries.

Then, the divergence field $\nabla \vec{c}$ is transformed to the frequency domain using the two-dimensional Discrete Cosine Transform (DCT-2):

$$B_{pq} = \frac{4}{MN} \sum_{m=0}^{M-1} \sum_{n=0}^{N-1} \nabla \vec{c}(m,n) \cos\left(\frac{\pi p(2m+1)}{2M}\right) \cos\left(\frac{\pi q(2n+1)}{2N}\right) \tag{2}$$

where $B_{pq}$ is the frequency-domain representation of the divergence field at position $(p, q)$; M and N are the grid dimensions in the $x$ and $y$ directions; $(m, n)$ represents the coordinates of a point in the spatial domain; $(p, q)$ indicates the position of a specific frequency component in the frequency-domain matrix.

Next, the Poisson equation is solved by applying DCT to transform the divergence $\nabla \vec{c}$ to the frequency domain:

$$A_{pq} = \frac{B_{pq}}{D_{pq}} \tag{3}$$

where:

$B_{pq}$ is the frequency-domain representation of the divergence field obtained via DCT-2.

$D_{pq}$ represents the discrete Laplacian operator in the frequency domain:

$$D_{pq}[i,j] = \frac{2\cos\left(\frac{\pi i}{N}\right) - 2}{dx^2} + \frac{2\cos\left(\frac{\pi j}{M}\right) - 2}{dy^2} \tag{4}$$

To avoid singularities, $D_{pq}[0,0]$ is set to 1, and the Neumann boundary condition is enforced by setting $A_{pq}[0,0] = 0$.

Finally, the potential field $a(x,y)$ is obtained by applying the Inverse Discrete Cosine Transform (IDCT):

$$a(x,y) = \sum_{p=0}^{M-1}\sum_{q=0}^{N-1} \alpha_p \alpha_q A_{pq} \cos\left(\frac{\pi p(2x+1)}{2M}\right) \cos\left(\frac{\pi q(2y+1)}{2N}\right) \tag{5}$$

$$\alpha_k = \begin{cases} \frac{1}{\sqrt{2}}, & if\ k = 0 \\ 1, & if\ k > 0 \end{cases} \tag{6}$$

where $\alpha_k$ is a normalization factor, $A_{pq}$ is the frequency-domain solution derived from the Poisson equation.

**Appendix 2: Stability evaluation of model results**

For individual-level SVM, each experiment includes 1,000 runs with randomized initial conditions, and each run of the algorithm performs up to 1,000 iterations to achieve a converged condition. The same experiment is repeated 50 times to assess the stability of the results. To evaluate these results, we consider three factors: goodness-of-fit directly measuring the quality of model fitting, similarity of results across different experiments reflecting result stability, and the results' closeness to the geographical coordinates of those locations (O/D points). A decision procedure is established as follows:

(1) Compute bidimensional regression among embeddings for the 50 experiments, obtaining a correlation matrix.

(2) Perform hierarchical clustering based on this correlation matrix to identify two stable groups of asymmetric patterns. The embedding group within the stable group with lower average stress is selected.

(3) Compute bidimensional regression between each embedding space and the two-dimensional geographic embeddings within the selected stable group. The geographic embeddings are generated by symmetric geographic distance through classical MDS. Embedding showing the best geographic proximity is selected as the final individual SVM result.

For local and global SVMs, the evaluation of model results is quite straightforward. Each experiment includes 100 runs with randomized initial conditions, and each run of algorithm performs up to 1,000 iterations. Also, 10 repetitions of the same experiment are conducted to select the results with the best goodness-of-fit.

**Appendix 3: The process of constructing the geographical space configuration**

We apply symmetric classical MDS to the geographical distance matrix of the states to obtain two-dimensional symmetric geographical space configurations. The symmetrical geographic distance between states is calculated by using Vincenty's formula (Vincenty 1975) to compute geodesic distances between latitude/longitude pairs on an ellipsoid. This approach is chosen instead of directly using geographical or projection coordinates to avoid significant dimensional discrepancies in the two-dimensional configurations, thereby facilitating analysis.

**Appendix 4: Symmetric–asymmetric decomposition of the normalized raw stress**

To enable structural attribution of model error without relying on the square root transformation inherent to classical stress metrics, we adopt a simplified definition of Normalized Raw Stress (NRS) as:

$$NRS = \frac{\sum (\delta_{ij} - d_{ij}(X,Y))^2}{\sum \delta_{ij}^2} \quad (7)$$

Building on this metric, we propose a symmetric–asymmetric decomposition of NRS to assess the contribution of different structural components to total model error.

Let the original dissimilarity matrix $\delta$ and the fitted embedded distance matrix $d$ be decomposed into symmetric and asymmetric components as:

$$\delta_{ij} = \delta_{ij}^s + \delta_{ij}^a, \quad d_{ij} = d_{ij}^s + d_{ij}^a \quad (8)$$

$$\delta_{ij}^s = \frac{\delta_{ij} + \delta_{ji}}{2}, \quad \delta_{ij}^a = \frac{\delta_{ij} - \delta_{ji}}{2} \quad (9)$$

$$d_{ij}^s = \frac{d_{ij} + d_{ji}}{2}, \quad d_{ij}^a = \frac{d_{ij} - d_{ji}}{2} \quad (10)$$

Defining the residual $e_{ij} = \delta_{ij} - d_{ij}$, the total residual sum of squares (stress) then can be expanded as:

$$\sum e_{ij}^2 = \sum (\delta_{ij}^s - d_{ij}^s)^2 + \sum (\delta_{ij}^a - d_{ij}^a)^2 + 2 \sum (\delta_{ij}^s - d_{ij}^s)(\delta_{ij}^a - d_{ij}^a) \quad (11)$$

The cross-term $\sum (\delta_{ij}^s - d_{ij}^s)(\delta_{ij}^a - d_{ij}^a)$ is negligible. Specifically, to justify the omission of the cross-term in Equation (11), we rely on the orthogonality of the symmetric and asymmetric components of a square matrix under the Frobenius inner product. Given the distance matrix $P$ and its transpose $P^T$ the symmetric part is defined as $S = \frac{P+P^T}{2}$, and the asymmetric part as $K = \frac{P-P^T}{2}$.

We show that these components are orthogonal under the Frobenius inner product:

$$\langle S, K \rangle_F = Tr(S^T K) = Tr(SK) = 0 \quad (12)$$

where $\langle \cdot,\cdot \rangle_F$ denotes the Frobenius inner product, $Tr(\cdot)$ denotes the trace operator.

To verify this, we expand:

$$SK = \left(\frac{P + P^T}{2}\right)\left(\frac{P - P^T}{2}\right) = \frac{1}{4}(P + P^T)(P - P^T) \quad (13)$$

Further simplifying:

$$(P + P^T)(P - P^T) = P^2 - PP^T + P^TP - (P^T)^2 \quad (14)$$

Taking the trace and using the properties $Tr(P^2) = Tr((P^T)^2)$ and $Tr(PP^T) = Tr(P^TP)$, we obtain:

$$\langle S, K \rangle_F = \frac{1}{4}[Tr(P^2) - Tr(PP^T) + Tr(PP^T) - Tr((P^T)^2)] = 0 \quad (15)$$

Thus, symmetric and skew-symmetric components are orthogonal in Frobenius space. By analogy, the symmetric and asymmetric residuals in our model can be considered orthogonal, allowing the cross-term in Equation (11) to be neglected.

Then, we can approximate the total residual sum of squares as:

$$RSS_{total} = RSS_{sym} + RSS_{asy} \quad (16)$$

We then denote the denominator of the $NRS$ formulation, $\sum \delta_{ij}^2$, as the Distance Sum of Squares (DSS), which captures the total energy of the embedded configuration. This quantity can be decomposed into contributions from the symmetric and asymmetric parts:

$$DSS_{sym} = \sum (\delta_{ij}^s)^2, \quad DSS_{sym} = \sum (\delta_{ij}^s)^2 \quad (17)$$

Then,

$$DSS_{total} = \sum \delta_{ij}^2 = \sum (\delta_{ij}^s + \delta_{ij}^a)^2 = \sum (\delta_{ij}^s)^2 + \sum (\delta_{ij}^a)^2 + 2\sum \delta_{ij}^s \cdot \delta_{ij}^a \quad (18)$$

Owing to the orthogonality between the symmetric component $\delta_{ij}^s$ and the asymmetric component $\delta_{ij}^a$, we can similarly decompose the numerator as:

$$DSS_{total} = DSS_{sym} + DSS_{asy} \tag{19}$$

The component-wise NRS are then defined as:

$$NRS_{sym} = \frac{RSS_{sym}}{DSS_{sym}}, \qquad NRS_{asy} = \frac{RSS_{asy}}{DSS_{asy}} \tag{20}$$

Letting:

$$w_{sym} = \frac{DSS_{sym}}{DSS_{total}}, \qquad w_{asy} = 1 - w_{sym} \tag{21}$$

The stress is then decomposed as:

$$NRS_{total} = \frac{RSS_{sym} + RSS_{asy}}{DSS_{total}} = w_{sym} \cdot NRS_{sym} + w_{asy} \cdot NRS_{asy} \tag{22}$$

**Appendix 5. Individual-level slide-vectors modeled based on data from 1965 to 2000**

**(1) 1965-1970**

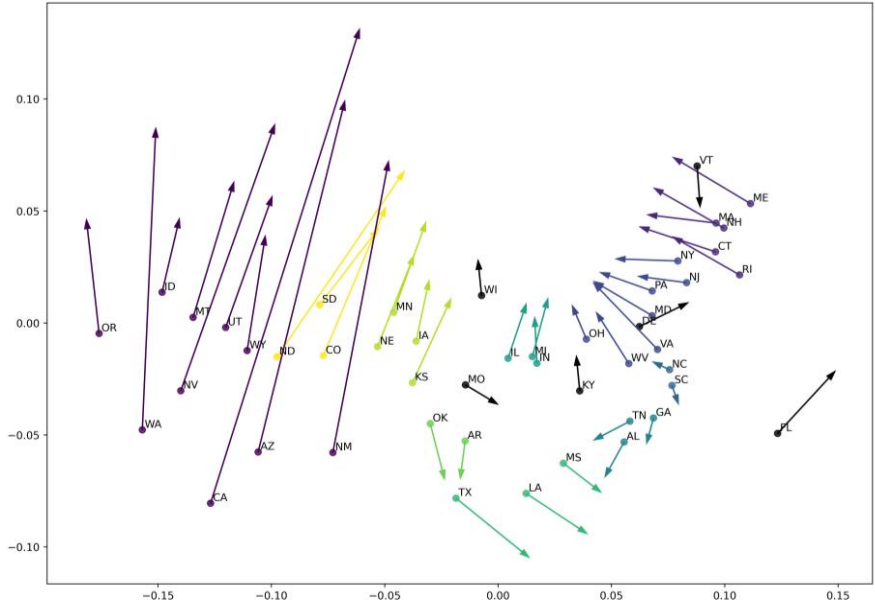

**(2) 1975-1980**

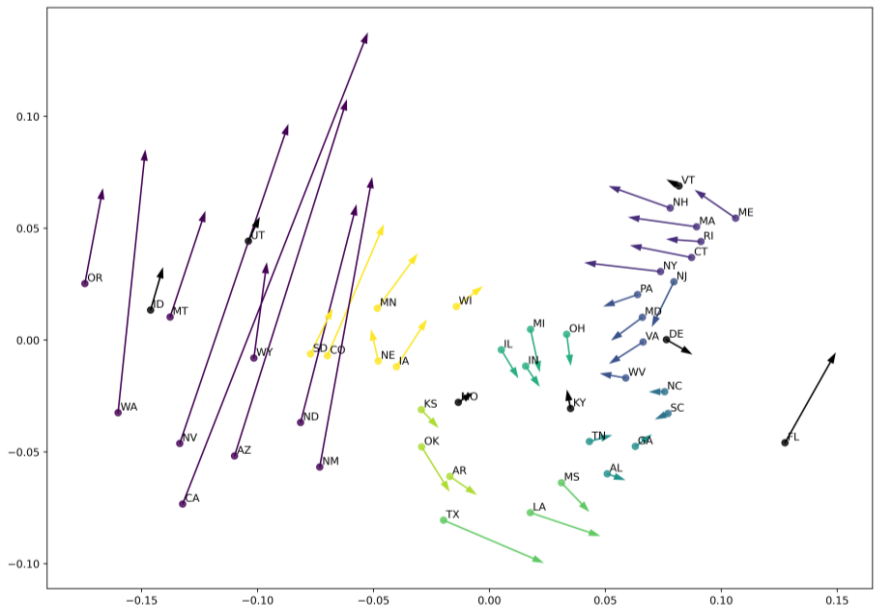

**(3) 1985-1990**

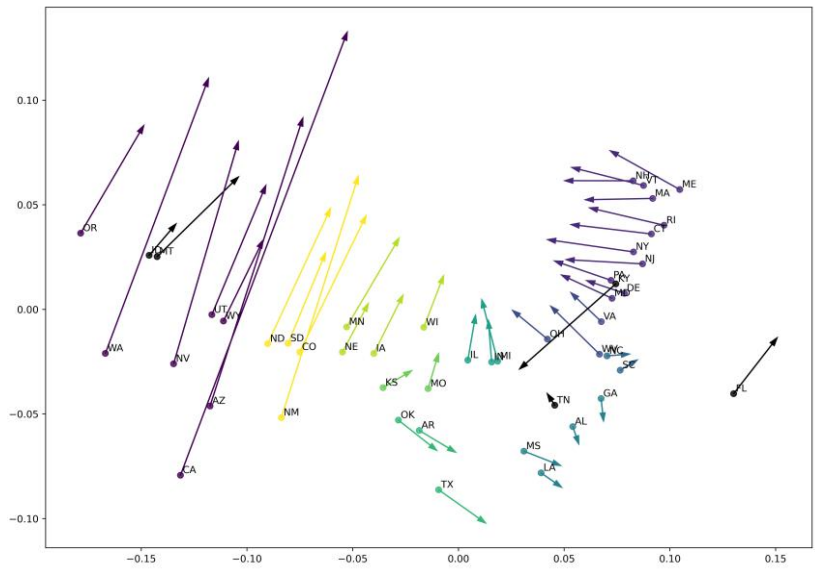

**(4) 1995-2000**

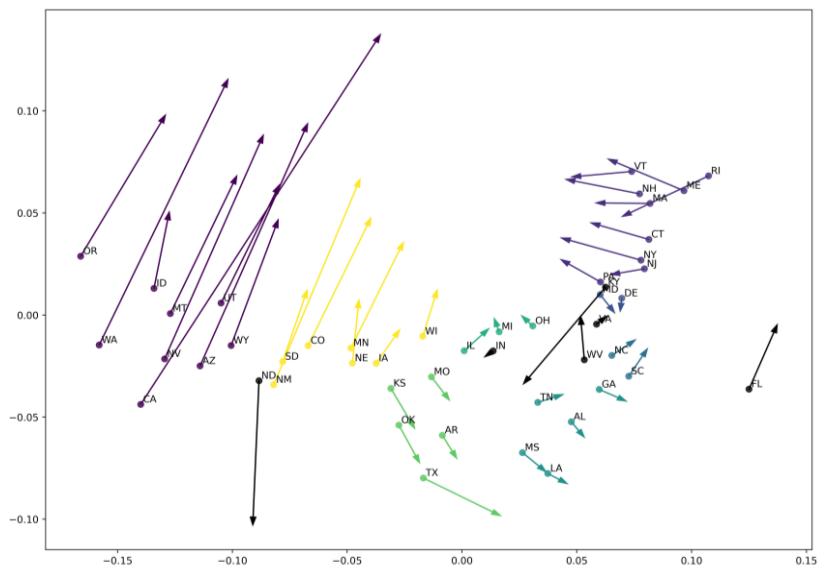

# Appendix 6: Delineation of typical watersheds

**(a) 1965-1970**

**(b) 1975-1980**

**(c) 1985-1990**

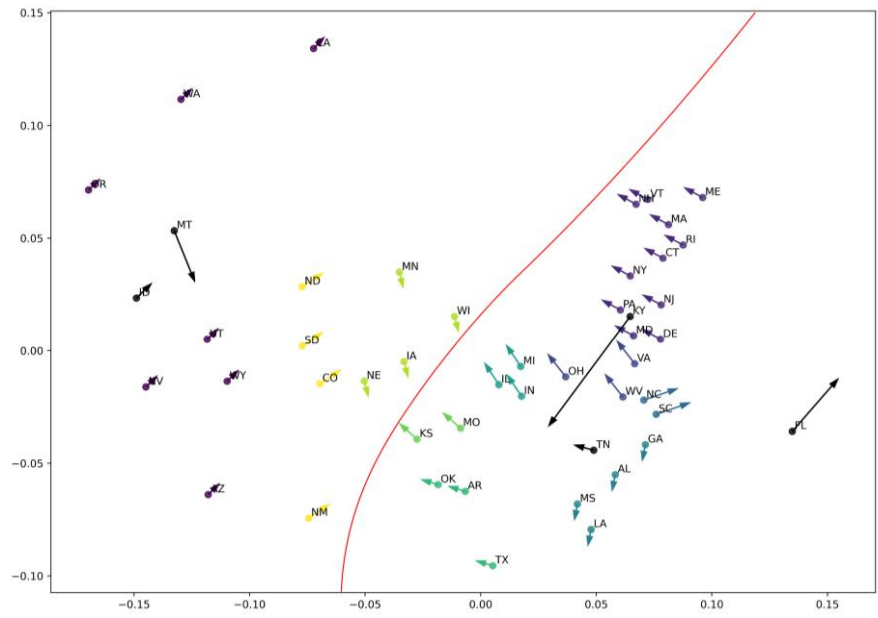

**Appendix 7: Details of DCI and DRC calculations**

To evaluate the goodness of asymmetry inference between local asymmetric regions based on the migration-scape, we devised a directional consistency index (DCI), which compares the inferred net migration resistance (difference in potential between two locations) to the real net migration resistance (difference between functional distances in opposite directions). DCI measures the consistency (1 as true and 0 as false) between the sign of the two net resistances between locations in the migration-scape. The real net migration resistance reflects the direction of the net migration flows between locations. For example, if the functional distance from A to B is smaller than that from B to A, the real net migration resistance from A to B is negative, meaning that traveling from A to B has lower resistance than traveling in the opposite direction. To be consistent with the real net migration resistance, the potential difference from A to B must be positive. Technically, we first selected two local asymmetry regions (either in one 'watershed' or in different 'watersheds'; second, we calculated the DCI between the two net migration resistance for all combinations of pair locations across the two regions; third, we calculated the proportion of matching pairs, whose DCI value is 1, as an aggregate indicator (DCR) for the goodness of asymmetry inference between the two regions.

**Reference:**
Borg, Ingwer, and Patrick JF Groenen. 2005. *Modern Multidimensional Scaling: Theory and Applications*. Springer Science & Business Media.
Borg, Ingwer, Patrick JF Groenen, and Patrick Mair. 2018. Applied Multidimensional Scaling and Unfolding. Springer.
Vincenty, Thaddeus. 1975. Direct and Inverse Solutions of Geodesics on the Ellipsoid with Application of Nested Equations. *Survey Review* 23 (176). Taylor & Francis:88–93.